\documentclass[twocolumn]{aastex701}

\usepackage{epsfig}
\usepackage{epstopdf}  
\usepackage{placeins}
\usepackage{graphicx,color}     
\usepackage{amssymb} 
\usepackage{float}
\usepackage{color}     
\usepackage{url}       
\usepackage{amsmath}
\usepackage{rotating}           
\usepackage{float}            
\usepackage{textcomp}           
\usepackage{dcolumn}
\usepackage{times}
\usepackage{natbib}
\usepackage{placeins}  
\usepackage{xcolor}
\usepackage{hyperref}
\usepackage{tabularx}
\usepackage[english]{babel}
\usepackage{amssymb}
\usepackage{pifont}
\usepackage{hyperref}
\usepackage[utf8]{inputenc}
\usepackage{bm}
\definecolor{ForestGreen}{RGB}{34,139,34}

\begin{document}

\shorttitle{Origin of Picoflare Jets}
\shortauthors{Bura et al.}
\title{On the Origin of Coronal Picoflare Jets}

\author[0009-0000-5018-9735]{Annu Bura}
\affiliation{Indian Institute of Astrophysics, Koramangala, Bangalore 560034, India; {\color{blue}{tanmoy.samanta@iiap.res.in}}}
\affiliation{Pondicherry University, R.V. Nagar, Kalapet 605014, Puducherry, India}
\email{annu.bura@iiap.res.in}  

\author[orcid=0000-0002-7788-6482]{Daniel Nóbrega-Siverio}
\affiliation{Instituto de Astrofísica de Canarias, E-38205 La Laguna, Tenerife, Spain}
\affiliation{Universidad de La Laguna, Dept. Astrofísica, E-38206 La Laguna, Tenerife, Spain}
\affiliation{Rosseland Centre for Solar Physics, University of Oslo, PO Box 1029 Blindern, 0315 Oslo, Norway}
\affiliation{Institute of Theoretical Astrophysics, University of Oslo, PO Box 1029 Blindern, 0315 Oslo, Norway}
\email{dnobres@gmail.com}

\author[orcid=0000-0002-9667-6392]{Tanmoy Samanta} 
\affiliation{Indian Institute of Astrophysics, Koramangala, Bangalore 560034, India; {\color{blue}{tanmoy.samanta@iiap.res.in}}}
\affiliation{Pondicherry University, R.V. Nagar, Kalapet 605014, Puducherry, India}
\email{tanmoy.samanta@iiap.res.in}

\author[orcid=0000-0003-0585-7030]{Jayant Joshi}
\affiliation{Indian Institute of Astrophysics, Koramangala, Bangalore 560034, India; {\color{blue}{tanmoy.samanta@iiap.res.in}}}
\email{jayant.joshi@iiap.res.in}

\begin{abstract}
Small-scale jet-like eruptions, such as picoflare jets and jetlets, are recognized as potential contributors to coronal heating and solar wind acceleration, yet their physical origin is still not fully established. Using ultra-high-resolution extreme ultraviolet imaging datasets from the Extreme Ultraviolet Imager on board the Solar Orbiter mission, we investigate tiny coronal jets observed off-limb in the Sun’s polar regions. Visual inspection reveals that the majority of these jets exhibit distinct morphological features, including a bright spire accompanied by a dark, eruptive jet component. We analyzed eleven of these jets in detail and found that their spatial and temporal scales are comparable to previously reported jetlets, while their kinetic energies are two to three orders of magnitude lower, placing them in the picoflare regime. The bright and dark components show distinct dynamics, with the dark structures generally displaying lower speeds. A comparison with coordinated Interface Region Imaging Spectrograph and the Atmospheric Imaging Assembly on board the Solar Dynamics Observatory data, together with 2.5D radiative-MHD simulations performed with the Bifrost code, reveals a one-to-one morphological correspondence between the dark counterparts and cool chromospheric surges accompanying the bright jet spire. This association suggests that flux emergence and magnetic reconnection at low atmospheric heights may produce coupled bright–dark structures, providing a plausible mechanism for the generation of picoflare jets. Our results demonstrate Solar Orbiter’s ability to resolve the dynamics of small-scale jets and place new constraints on their origin.
\end{abstract}

\section{Introduction}
Solar coronal jets are collimated plasma ejections from the solar atmosphere and are observed across a wide range of wavelengths (for a review of jets, see \citealt{2016SSRv..201....1R, 2021RSPSA.47700217S, 2022FrASS...920183S}), from X-rays (\citealt{1992PASJ...44L.173S, 1996ApJ...464.1016C, 1996PASJ...48..123S, 2010ApJ...720..757M, 2015Natur.523..437S, 2022ApJ...940...85S}) to Extreme Ultraviolet (EUV) (\citealt{1997Natur.386..811I, 2008ApJ...673L.211M, 2009SoPh..259...87N, 2010A&A...519A..49H, 2014Sci...346A.315T, 2016ApJ...828L...9S, 2020A&A...639A..22J, 2025A&A...702A.188N}) and sometimes exhibit both cool and hot plasma components.\\
Beyond the large-scale coronal jets, the solar atmosphere hosts numerous jet-like phenomena in the transition region and chromosphere at even smaller spatial and temporal scales, such as transition region network jets \citep{2014Sci...346A.315T, 2016SoPh..291.1129N, 2018ApJ...854...92T, 2022A&A...660A.116G}, chromospheric jets \citep{2011A&A...533A..76K, 2019ApJ...870...90B, 2024A&A...691A.198J, 2025A&A...698A.174B, 2025ApJ...985L..47B} and spicules \citep{2007PASJ...59S.655D, 2012ApJ...759...18P, 2014ApJ...792L..15P, 2015ApJ...815L..16S, 2019Sci...366..890S, 2021A&A...647A.147B, 2023ApJ...944..171B}. These small-scale eruptions, which are more frequent and shorter-lived than coronal jets, provide a potential mechanism for supplying plasma and energy to the outer corona and solar wind. (\citealt{2019Sci...366..890S, 2021RSPSA.47700217S, 2023Sci...381..867C, 2023ApJ...945...28R, 2025ApJ...983L...7B}).\\
\FloatBarrier
\begin{deluxetable*}{cllcccccccc}
\tablenum{1}
\tablecaption{Summary of the observations analyzed in this study.\label{tab:obs}}
\tablewidth{0pt}
\tablehead{
\colhead{Dataset} &
\colhead{Date} &
\colhead{\shortstack{Time \\ (UT)}} &
\colhead{Instrument} &
\colhead{Location} &
\colhead{\shortstack{Distance \\ (AU)}} &
\colhead{Cadence} &
\colhead{\shortstack{Pixel scale \\ ($''$)}} &
\colhead{\shortstack{Resolved km \\ on the Sun}} &
\colhead{\shortstack{No. of jets \\ analyzed}}
}
\startdata
D1 & 2021-09-14 & 03:19--04:18 & HRI$_{\mathrm{EUV}}$ & North pole & 0.587 & 20\,s & 0.492 & 420 & 1 (Jet~1) \\
D2 & 2021-09-14 & 04:35--05:50 & HRI$_{\mathrm{EUV}}$ & South pole & 0.587 & 20\,s & 0.492 & 420 & 2 (Jet~2, Jet~3) \\
   &             & 05:53--06:11 &                     &            &       & 5\,s  &       &       &   \\
D3 & 2022-03-30 & 04:30--04:59 & HRI$_{\mathrm{EUV}}$ & South pole & 0.332 & 3\,s  & 0.492 & 240 & 6 (Jet~4 -- Jet~9) \\
D4 & 2024-04-05 & 19:59--23:34 & HRI$_{\mathrm{EUV}}$ & West limb  & 0.295 & 16\,s & 0.492 & 210 & 1 (Jet~10) \\
D5 & 2019-08-07 & 22:00--23:00 & IRIS                & North pole & 1.0   & 74\,s (Raster) & 0.33 & 480 & 1 (Jet~11) \\
   &             &              & AIA                 &            &       & 12\,s & 0.6   & 870 &   \\
\enddata
\end{deluxetable*}

In coronal jets, we often observed the cool component as a surge or filament eruption in H$\alpha$ or EUV images, whereas the hot component observed as bright EUV or X-ray spires (\citealt{1999ApJ...513L..75C, 2013ApJ...769..134M, 2009SoPh..255...79C, 2017A&A...606A...4M, 2017ApJ...851...67S, 2020ApJ...889..187S}). Surges, for instance, are chromospheric ejections that reach coronal heights and are commonly observed in H$\alpha$ and other chromospheric lines as dark, elongated structures (\citealt{1973SoPh...32..139R, 1984SoPh...94..133S, 2007A&A...469..331J, 2012ApJ...752...70U, 2018ApJ...854...92T, 2024A&A...686A.218N}). They are most often associated with emerging flux regions and are frequently accompanied by hot EUV or X-ray jets, indicating that magnetic reconnection between newly emerging and ambient fields is a primary driver (\citealt{1992PASJ...44L.173S, 1996PASJ...48..353Y, 2014ApJ...794..140V, 2013ApJ...771...20M, 2016ApJ...822...18N}). Filament eruptions, particularly at small scales (often termed `minifilament eruptions', \citealt{2015Natur.523..437S}), are usually linked to flux cancellation events, and observed in EUV images from Atmospheric Imaging Assembly (AIA; \citealt{2012SoPh..275...17L}) on board Solar Dynamics Observatory (SDO; \citealt{2012SoPh..275....3P}) as absorption features, especially in the \ion{He}{2} 304 \AA\ channel. The destabilization and eruption of these minifilaments drives the jet spire, while magnetic reconnection beneath the erupting structure produces the associated jet bright point (JBP) (\citealt{2011ApJ...738L..20H, 2012ApJ...745..164S, 2014ApJ...783...11A, 2015Natur.523..437S, 2016ApJ...821..100S, 2017Natur.544..452W, 2019ApJ...873...93K, 2020ApJ...896L..18S}). These observations of coexisting bright and dark components in coronal jets motivate the investigation of similar features at smaller spatial and temporal scales, where tiny jet-like eruptions may provide crucial insight into the physical processes driving these multi-thermal structures.\\
Many numerical simulations have also investigated the ejection of hot (bright spire) and cool (dark surge) plasma associated with magnetic flux emergence through the low atmosphere into the corona (\citealt{1996PASJ...48..353Y, 2012ApJ...751..152J, 2013ApJ...771...20M, 2016ApJ...822...18N, 2018ApJ...852...16Y, 2023ApJ...947L..17L}). While such models reproduce surge-like ejections successfully, realistic radiative-MHD simulations of minifilament eruptions are still lacking. Therefore, to directly validate these processes, we require observations capable of resolving the intricate dynamics of both bright and dark plasma structures in jet evolution.\\
The Extreme Ultraviolet Imager (EUI; \citealt{2020A&A...642A...8R}) on board Solar Orbiter (\citealt{2020A&A...642A...1M}) provides such capability, offering unprecedented spatial resolution and high cadence. With these advances, EUI has revealed numerous small-scale transient events, such as campfires (\citealt{2021A&A...656L...4B, 2021ApJ...921L..20P, 2022A&A...660A.143K, 2025A&A...699A.138N}), tiny inverted Y-shaped coronal jets with energies in the nanoflare and picoflare ranges (\citealt{2022A&A...664A..28M, 2023ApJ...943...24P, 2023Sci...381..867C, 2025A&A...694A..71C, 2024A&A...686A.279S}), and nanojets (\citealt{2025ApJ...985L..12G, 2025ApJ...988L..65B}). These studies highlight the potential role of small-scale jet-like eruptions, such as jetlets (\citealt{2023ApJ...945...28R, 2024ApJ...963....4S}) and picoflare jets (\citealt{2023Sci...381..867C, 2025A&A...694A..71C}), in contributing significantly to the mass and energy flux of the solar wind. Among these, picoflare jets represent the smallest resolved reconnection-driven jets observed to date, and they have been suggested as potential contributors to coronal heating and solar wind acceleration. They have widths of $\sim$100~km, evolve on timescales of 20-100\,s, and often show inverted Y-shaped structures. Yet, their physical origin remains elusive.\\
\begin{figure*}[t]
\centering
\includegraphics[width=0.95\textwidth]{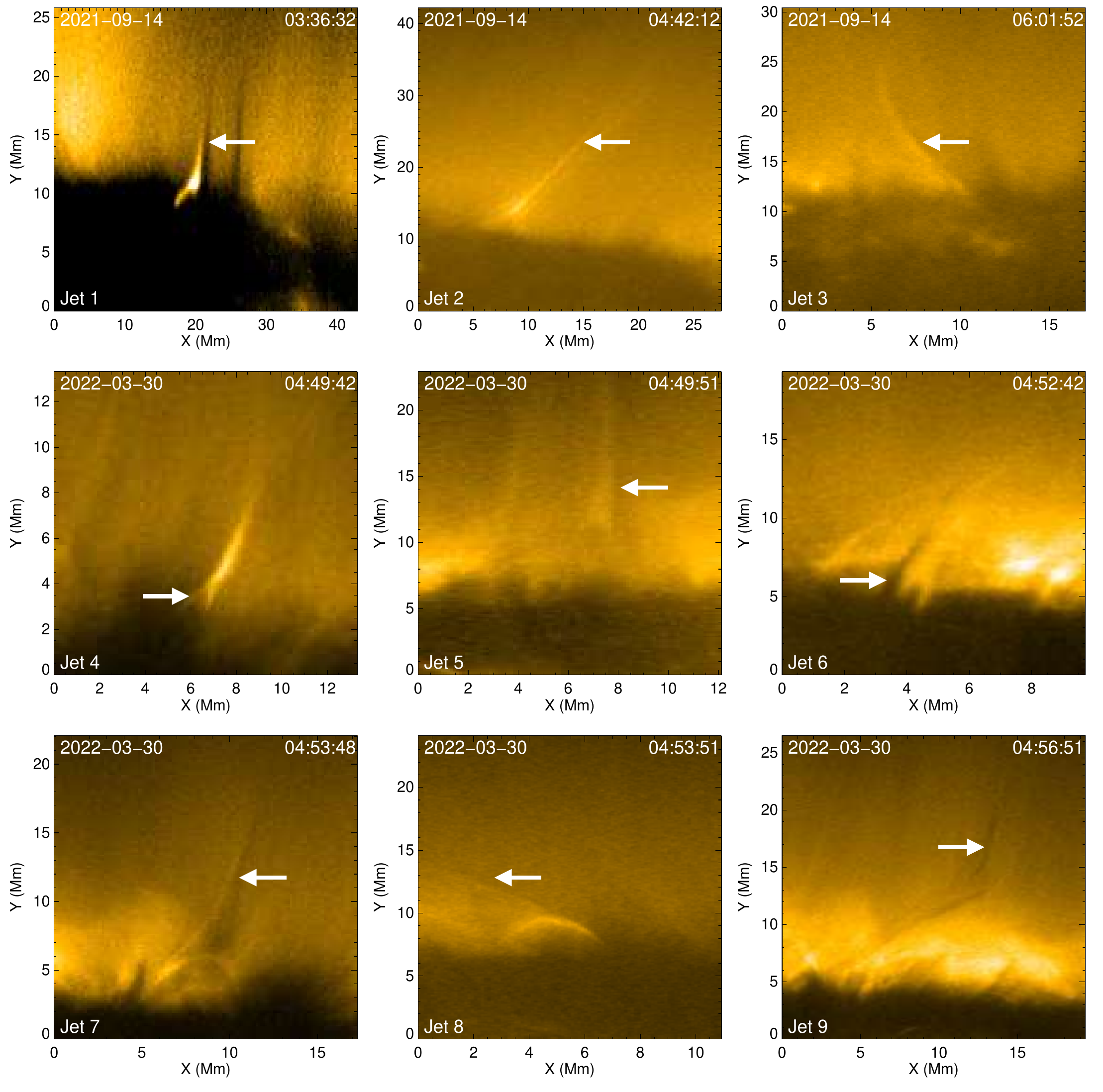}
\caption{Examples of jets observed off-limb in the Sun's polar regions using HRI$_{\mathrm{EUV}}$ on board Solar Orbiter with a passband centered at 174\,\AA, each accompanied by a narrow, collimated dark structure indicated by white arrows. These jets are selected from three datasets: the north pole observation on September 14, 2021 (Jet 1), the south pole observations on September 14, 2021 (Jets 2-3), and on March 30, 2022 (Jets 4-9). An animation of this figure is available online. The real-time duration of the animation is 10~s.}
\label{fig:1}
\end{figure*}
\begin{figure*}[t]
\centering
\includegraphics[width=\textwidth]{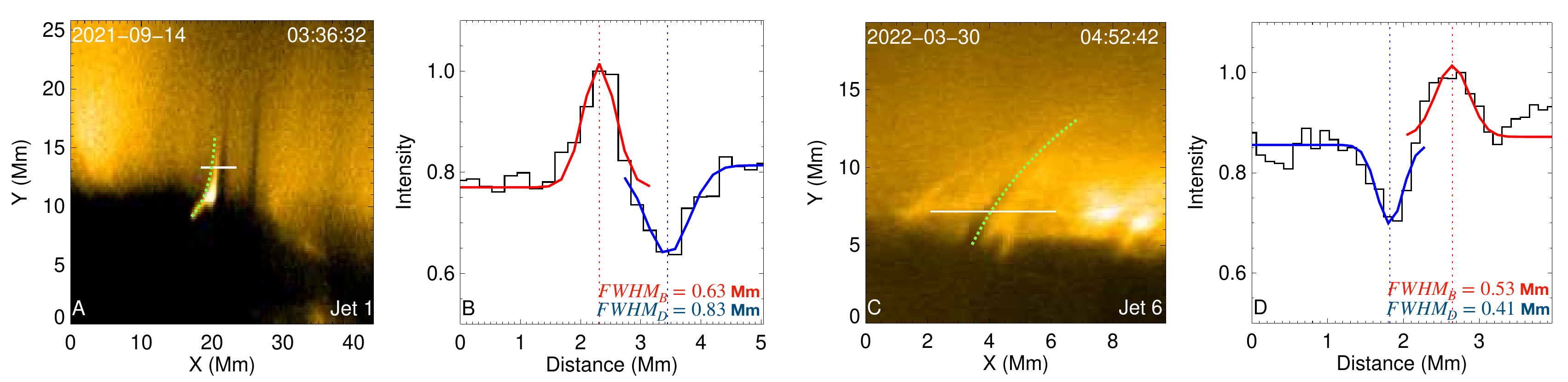}
\caption{Width estimation for jets observed in HRI$_{\mathrm{EUV}}$. Panel (A) shows the observation of Jet 1 in HRI$_{\mathrm{EUV}}$ with a white artificial slit marking the location used to extract the intensity profile. Panel (B) shows the corresponding intensity profile (histogram mode), where two Gaussian functions are fitted to the peak (red curve) and dip (blue curve) in the profile, corresponding to the adjacent bright and dark structures, to estimate their FWHM. Panels (C) and (D) show a similar analysis for jet 6. The green dashed curve in panels (A) and (C) represents the spline-interpolated curve from the jet base at the visible surface to the top of the spire, used to estimate the jet lengths.}
\label{fig:2}
\end{figure*}
\begin{figure*}[t]
\centering
\includegraphics[width=\textwidth]{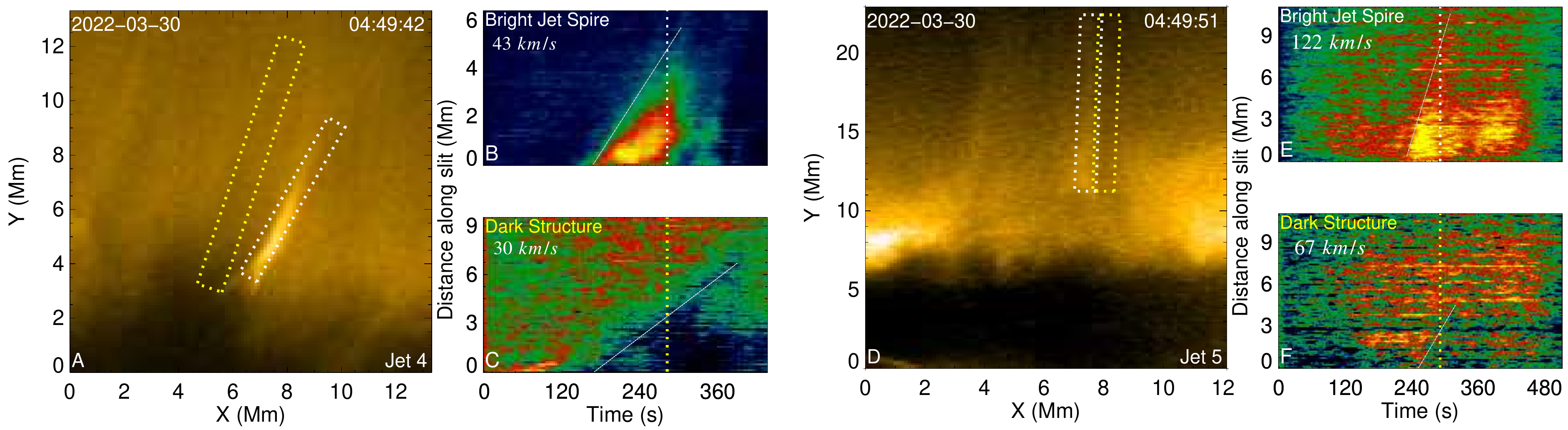}
\caption{Speed estimation of jets observed in HRI$_{\mathrm{EUV}}$. Panel (A) shows Jet 4 in HRI$_{\mathrm{EUV}}$ with two artificial slits overlaid: a white rectangle (5 pixels wide) and a yellow rectangle (7 pixels wide), used to construct space-time diagrams for the bright jet spire and the associated dark structure, respectively. Panels (B) and (C) display the resulting space-time plots along these artificial slits. For the bright jet spire, the maximum intensity values along the white slit are used to construct the space–time diagram, and the propagation speed is obtained by tracking the inclined intensity ridge corresponding to the upper part of the spire. For the dark structure, the minimum intensity values along the yellow slit are used in an analogous manner. Panels (D)–(F) present the same analysis for Jet 5.}
\label{fig:3}
\end{figure*}
In this study, we combine high-resolution EUI observations with numerical simulations to investigate tiny coronal jets observed off-limb in the Sun’s polar regions. By leveraging our unique\
observations, which reveal both bright and dark components at very small spatial scales, we aim to shed light on the origin of picoflare jets.
\section{Data Analysis, Results and Discussion}
\subsection{Observations and Data Reduction}
We investigate the bright and dark structures of tiny coronal jets observed off-limb in the Sun’s polar regions. While such coupled bright and dark structures are commonly observed in larger coronal jets, in our work, we investigate similar features at smaller spatial and temporal scales using High Resolution Imager (HRI) data on board Solar Orbiter, with the aim of gaining insights into their underlying physical processes. For this purpose, we reviewed the publicly available HRI$_{\mathrm{EUV}}$ animation datasets from Solar Orbiter, focusing on off-limb observations. Through the visual inspection of these observations, we noticed that the majority of jets display the simultaneous presence of a bright spire and a narrow, collimated dark structure. From these, we selected ten well-isolated jets from four datasets in which the bright and dark structures could be clearly distinguished and whose evolution could be followed. In total, we analyzed eleven coronal jets using these four HRI$_{\mathrm{EUV}}$ datasets, along with one coordinated observation from the Interface Region Imaging Spectrograph (IRIS; \citealt{2014SoPh..289.2733D}) and AIA on board SDO. Our analysis is focused on the HRI$_{\mathrm{EUV}}$ channel centered at 174 \AA, which is mainly sensitive to plasma at temperatures of $\sim$1.0~MK. \\
Table~\ref{tab:obs} provides a comprehensive summary of the observations analyzed in this study. During the first dataset, acquired on September 14, 2021 (hereafter D1), Solar Orbiter was located at a heliocentric distance of $\sim$0.587~AU.
HRI$_{\text{EUV}}$ observed the north solar pole from 03:19~UT to 04:18~UT with a cadence of 20\,s and a pixel scale of 0.492$''$, corresponding to a pixel size of $\sim$0.210~Mm. The second observation (hereafter D2), also on September 14, 2021, covered the south polar region from 04:35~UT to 05:50~UT at a cadence of 20~s, and from 05:53~UT to 06:11~UT at a higher cadence of 5~s, with the same pixel scale of $\sim$0.210~Mm. The third observation (hereafter D3), conducted on March 30, 2022, when Solar Orbiter was at $\sim$0.332~AU, observed the south pole from 04:30~UT to 04:59~UT. This dataset provided enhanced temporal and spatial resolution with a cadence of 3\,s and a pixel size of $\sim$0.120~Mm. The fourth observation at west limb (hereafter D4), taken on April 5, 2024, at a heliocentric distance of $\sim$0.295~AU, covered the time interval from 19:59~UT to 23:34~UT, with a cadence of 16\,s and a pixel size of $\sim$0.105~Mm. We analyzed one jet from D1, two jets from D2, six jets from D3, and one jet from D4.\\
The full width at half maximum (FWHM) of the core of the point spread function of HRI$_{\mathrm{EUV}}$ is about two pixels \citep{2021A&A...656L...4B, 2023Sci...381..867C}, corresponding to a spatial resolution of 0.984$''$ which is $\sim$0.420~Mm for D1 and D2, $\sim$0.240~Mm for D3 and $\sim$0.210~Mm for D4. We used level 2 datasets of HRI$_{\text{EUV}}$ from the latest EUI data release 6.0 \citep{EUI_DR6}. To mitigate the effects of residual spacecraft jitter in the data, we followed the approach described in \citet{2022A&A...667A.166C}, also employed in earlier HRI$_{\text{EUV}}$ observations (\citealt{2025ApJ...988L..65B}).\\
With the aim of providing a complementary analysis, in the fifth dataset (hereafter D5), we analyzed coordinated observations from IRIS and AIA, acquired on August\,7,\,2019. For this observation, IRIS performed a 64-step raster scan covering a field of view (FOV) of $22'' \times 119''$. The step size and raster cadence were 1.2\,s and 74\,s, respectively. The observation, targeting a coronal hole at the north pole, was conducted from 22:00~UT to 23:00~UT, with IRIS pointing centered at ($-$0.715$''$, 828.867$''$). The plate scale was 0.3327$''$ per pixel. IRIS provided slit-jaw images (SJIs) in 2832 \AA\ passband at a cadence of 9\,s, which is mainly sensitive to the plasma emission from the upper photosphere at a temperature of T\,$\sim$\,10$^{3.8}$\,K. We used Level 2 data for our analysis. For the AIA data, we used the 171 \AA\ passband, which is sensitive to emission from \ion{Fe}{9}, with a characteristic formation temperature of T\,$\sim$\,10$^{5.8}$\,K. We calibrated AIA data using the \texttt{aia\_prep.pro} routine available in SolarSoft. To co-align the IRIS and AIA data, we matched photospheric features in images from both IRIS SJI 2832 \AA\ and AIA 1700 \AA\ passband.
\subsection{Numerical Experiment}
\label{s-NE}
To provide support to our observations, we use a radiative-magnetohydrodynamic (radiative-MHD) numerical simulation performed with the Bifrost code \citep{2011A&A...531A.154G}. The experiment, presented and detailed described in \citet{2017ApJ...850..153N, 2018ApJ...858....8N}, is a 2.5D simulation of magnetic flux emergence in a coronal hole that leads to the formation of a coronal jet along with a surge. The realism of the simulation has been proven valuable for understanding the transition-region emission of surges observed with IRIS. In addition, it has been used to explain plasmoid signatures in chromospheric and transition region lines (\citealt{2017ApJ...851L...6R}).  As shown in the following, it also shares resemblance with the observed jets and dark structures presented in this study. 
\subsection{Dynamical Properties of Bright and Dark Structures}
Figure~\ref{fig:1} shows jets observed at the solar poles using HRI$_{\text{EUV}}$ on board Solar Orbiter in the 174 \AA\ passband. Each jet exhibits a bright spire and a collimated dark structure (indicated by white arrows) and has the shape of an inverted `Y' (or '$\lambda$') shape (\citealt{1994ApJ...431L..51S, 1995Natur.375...42Y}). These bright and dark structures are the primary focus of the analysis presented in this study.\\
We analyzed the morphological and dynamical properties of both the bright jet spires and their associated dark structures. To estimate the width of these structures, we placed an artificial slit (shown as a white solid line in panels (A) and (C) of Figure~\ref{fig:2}) across the jet, intersecting both the bright and dark structures. We employed the same method for all of the jets, and for illustrative purposes, we have shown only cases 1 and 6. To maintain a uniform orientation across all jets and facilitate visual interpretation, we mirrored the images of Jet 1 about the Y-axis, bringing the structure of the jet in line with the typical jet morphology commonly seen in the literature associated with jets \citep{1992PASJ...44L.173S, 1996ApJ...464.1016C}. The corresponding intensity profiles along these slit are shown as histograms in panels (B) and (D). To quantify the width, we fitted two separate Gaussian functions: one to the intensity peak representing the bright jet spire (red curve), and the other to the intensity dip corresponding to the dark structure (blue curve). The FWHM of these Gaussian fits provides a measure of their width. This analysis was applied consistently across all jets in the dataset. We found that the width of the bright spire ranges from 0.35 to 1.75~Mm with a mean value of 0.69~Mm, while the width of the dark structure ranges from 0.23 to 0.83~Mm with a mean value of 0.49~Mm. We also estimated the projected length of these jets from the jet base at the visible surface to the top of the spire using a cubic spline interpolation method. For this analysis, we visually selected four points along the jet axis and interpolated them to 100 points using the IDL routine spline.pro to obtain a smooth curve. We repeated this procedure five times to reduce bias from manual selection and took the average as the final jet length. The resulting spline-interpolated curve is shown as a green dashed curve in panels (A) and (C) for Jet 1 and Jet 6, respectively. From this analysis, we found that the jet lengths range from approximately 7.85 to 29.74~Mm, with a mean length of 16.90~Mm. We estimated the uncertainties in the measured jet lengths from repeated spline measurements and found them to be very small relative to the jet lengths. For example, for Jet 1 shown in panel (A) of Figure 2, the
uncertainty was $\sim$0.12~Mm.\\
\begin{figure*}[t]
\centering
\includegraphics[width=0.9\textwidth]{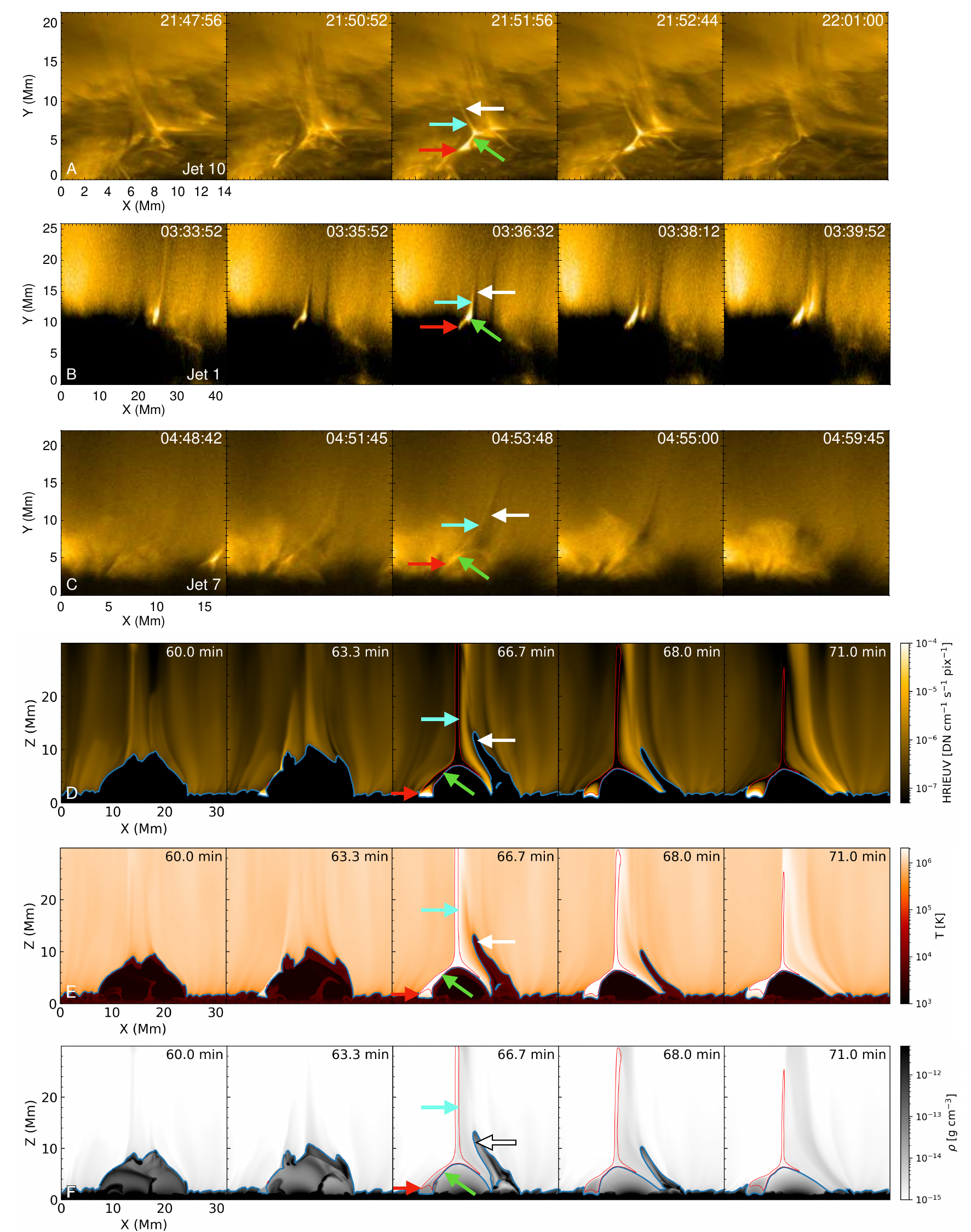}
\caption{Temporal evolution of jets and comparison with simulation. Panel (A): shows the temporal evolution of Jet 10 observed on April 5, 2024. Panel (B): shows the temporal evolution of Jet 1 observed on September 14, 2021, during the north polar observation. Panel (C): shows the temporal evolution of Jet 7 observed on March 30, 2022. Panel (D): Comparison with a flux emergence simulation from \citealt{2018ApJ...858....8N},
showing synthetic emission in the HRI$_{\mathrm{EUV}}$ 174 \AA\ filter. Blue and red contours correspond to plasma at T\,=\,0.1\,MK and T\,=\,2\,MK, highlighting the locations of the cool and hot ejections, respectively. The red, green, cyan, and white arrows mark the JBP, current sheet, bright spire, and dark structure, respectively. Panels (E–F) show the temporal evolution of the temperature and density maps of the jets, with the same temperature contours from Panel D overlaid. Note that the 2.5D synthetic 174 \AA\ image appears dark in regions where plasma temperatures lie outside the passband’s sensitivity, such as the $\sim$2 MK spire and $\sim$0.1 MK dark jet; however, the temperature maps clearly reveal them. The animation of panels (D)-(F) is available online. The real-time duration of the animation is 13~s.}
\label{fig:4}
\end{figure*}
To estimate the propagation speeds of both the bright jet spires and the associated dark structures, we constructed space–time diagrams by placing two separate artificial slits along the direction of jet propagation, one along the bright jet spire (indicated by a white dashed rectangular box) and the other along the dark structure (yellow dashed rectangular box), as shown in panels (A) and (D) of Figure~\ref{fig:3} for Jets 4 and 5. Within these slits, the maximum intensity values were extracted for the bright spire (white slit) at each timestamp, while the minimum intensity values were extracted for the dark structure (yellow slit) to construct their respective space–time diagrams. The corresponding space-time diagrams for the bright spires are displayed in panels (B) and (E), while those for the dark structures are shown in panels (C) and (F) of Figure~\ref{fig:3}. In the space-time diagrams, the motion of the upper parts of both structures appears as inclined ridges. By manually tracking these ridges and measuring their slopes, we estimated their projected plane-of-sky speeds. From this space–time analysis, we find that the bright jet spires have speeds ranging from 43 to 132~km\,s$^{-1}$ (mean:\,85~km\,s$^{-1}$), while the dark structures propagate more slowly, with speeds between 30 and 67~km\,s$^{-1}$ (mean:\,49~km\,s$^{-1}$). These measured jet speed represents the motion of the upper part of the bright spire rather than the bulk plasma flow, as latter is often ambiguous to track due to mixed flows. However, we find that the uncertainties in the measured speeds, estimated from five repeated measurements of the ridge positions, are very small and therefore unlikely to significantly affect the derived kinetic energy of the bright spire. For example, in the case of Jet 4, shown in panel (A) of Figure~3, the uncertainties were 2.7 km s$^{-1}$ and 1.3 km s$^{-1}$ for the bright and dark structures, respectively. We note, however, that these uncertainty values should be considered as lower limits, as projection effects also complicate the estimation of the speed and its associated uncertainties. These estimated speed values are consistent with previously reported speeds for chromospheric ejections, such as surges, based on both observations 
\citep{1984SoPh...94..133S, 1999ApJ...513L..75C, 2013SoPh..288...39Y, 2020A&A...639A..19V, 2024A&A...686A.218N} and numerical simulations \citep{2013ApJ...771...20M, 2015A&A...576A...4M, 2016ApJ...822...18N, 2023ApJ...947L..17L}. In all cases, the bright jet spire moves faster than the associated dark structures. The lifetimes of the jets were determined by visually tracking the duration over which they remained discernible in the time series. Their lifetime ranges from 3.4 to 11.5 minutes, with an average lifetime of approximately 6.2 minutes. The measured properties of the observed jets are summarized in Table~\ref{tab:jet_props}.\\
We find that the spatial and temporal scales of our jets lie between those of picoflare jets and jetlets. \citet{2023Sci...381..867C} reported that picoflare jets are a few hundred km wide, have speeds of order \(100\ \mathrm{km\,s^{-1}}\), and carry kinetic energies in the picoflare range. Our events show bright-spire widths of 0.35--1.75~Mm, lengths of 7.85--29.74~Mm, lifetimes of 3.4--11.5~minutes, and spire speeds of 43--132~km\,s\(^{-1}\). These size and lifetime values are also consistent with reported jetlet properties (widths: 0.6--3~Mm, lengths: 9--27~Mm, lifetimes up to several minutes;  \citealt{2019ApJ...887L...8P,2023ApJ...945...28R}), while the propagation speeds are comparable to picoflare jets. However, \citet{2023Sci...381..867C} did not report any associated dark structures in their study. One reason could be that most of their jets were observed on disk, where strong background emission makes it difficult to detect faint absorption features. By contrast, our jets are observed off-limb at the Sun's polar regions, where the reduced background brightness allows dark structures to be seen more clearly. Another possibility is that their jets are smaller in spatial and temporal scales compared to ours, which makes it harder to identify any associated dark counterparts. Moreover, we require enough dense and cool material to absorb the EUV emission from the background, which is primarily due to the presence of neutral hydrogen and helium (\ion{He}{1} and \ion{He}{2}) (\citealt{2005ApJ...622..714A}), otherwise dark structures may remain too transparent to be observed.\\
\FloatBarrier
\begin{deluxetable*}{cccccccc}
\tablenum{2}
\tablecaption{Measured properties of the observed jets in HRI$_{\mathrm{EUV}}$ on board Solar Orbiter \label{tab:jet_props}}
\tablewidth{0pt}
\tablehead{
\colhead{Jet No.} & 
\colhead{\shortstack{Bright spire \\ width (Mm)}} & 
\colhead{\shortstack{Dark structure \\ width (Mm)}} & 
\colhead{\shortstack{Bright spire \\ speed (km s$^{-1}$)}} & 
\colhead{\shortstack{Dark structure \\ speed (km s$^{-1}$)}} & 
\colhead{Length (Mm)} & 
\colhead{\shortstack{Lifetime \\ (minutes)}} & 
\colhead{\shortstack{Bright spire \\ Kinetic Energy \\(erg)}}
}
\startdata
1  & 0.63 & 0.83 & 65   & 55   &  8.67 &  5.00 & $1.91 \times 10^{22}$ \\
2  & 1.75 & $-$   & $-$  & $-$  & 29.74 &  7.00 & $-$                \\
3  & 0.89 & 0.77 & 63  & $-$  & 24.18 &  3.40 & $9.98 \times 10^{22}$ \\
4  & 0.63 & 0.67 & 43  & 30   &  7.58 &  5.35 & $7.30 \times 10^{21}$ \\
5  & 0.47 & 0.23 & 122  & 67   & 21.09 &  4.55 & $9.10 \times 10^{22}$ \\
6  & 0.53 & 0.41 & 63   & 39   & 13.05 &  4.90 & $1.91 \times 10^{22}$ \\
7  & 0.37 & 0.45 & 106  & 65   & 15.65 & 10.15 & $3.16 \times 10^{22}$ \\
8  & 0.63 & 0.28 & 63   & $-$  & 13.31 &  4.45 & $2.75 \times 10^{22}$ \\
9  & $-$  & $-$  & 132  & 47   & 23.74 &  5.70 & $-$                \\
10 & 0.35 & 0.28 & 110  & 39   & 12.03 & 11.46 & $2.34 \times 10^{22}$ \\
\enddata
\tablecomments{Dashes indicate where properties could not be measured reliably.}
\end{deluxetable*}
The properties of our jets are also comparable to the thin jets associated with coronal bright points, as reported in a recent study by \citet{2025A&A...702A.188N}. However, no dark structures were detected in those CBP-related jets, likely due to differences in the driving mechanisms. Furthermore, in comparison to spicules, which have widths of 0.2--0.5~Mm, lengths of 5--10~Mm, a lifetime of 5-15 minutes, and rise with speeds of 15--150~km~s$^{-1}$ \citep{2007PASJ...59S.655D, 2012ApJ...759...18P, 2015ApJ...815L..16S}, our jets occupy a comparatively larger spatio-temporal scale. Interestingly, \citet{2019Sci...366..890S} reported that enhanced spicular activity can channel hot plasma into the corona and heat the upper solar atmosphere. In their on-disk observations, spicules manifest as dark features in H$\alpha$, with coronal emission in AIA 171~\AA\ generally appearing at their tops. An open question that arises is whether spicules share the same underlying origin as the coupled bright–dark structures revealed in our EUI jets. We expect that future high-resolution, multi-wavelength observations will provide the critical evidence needed to establish or refute this connection.
\begin{figure*}[t]
\centering
\includegraphics[width=\textwidth]{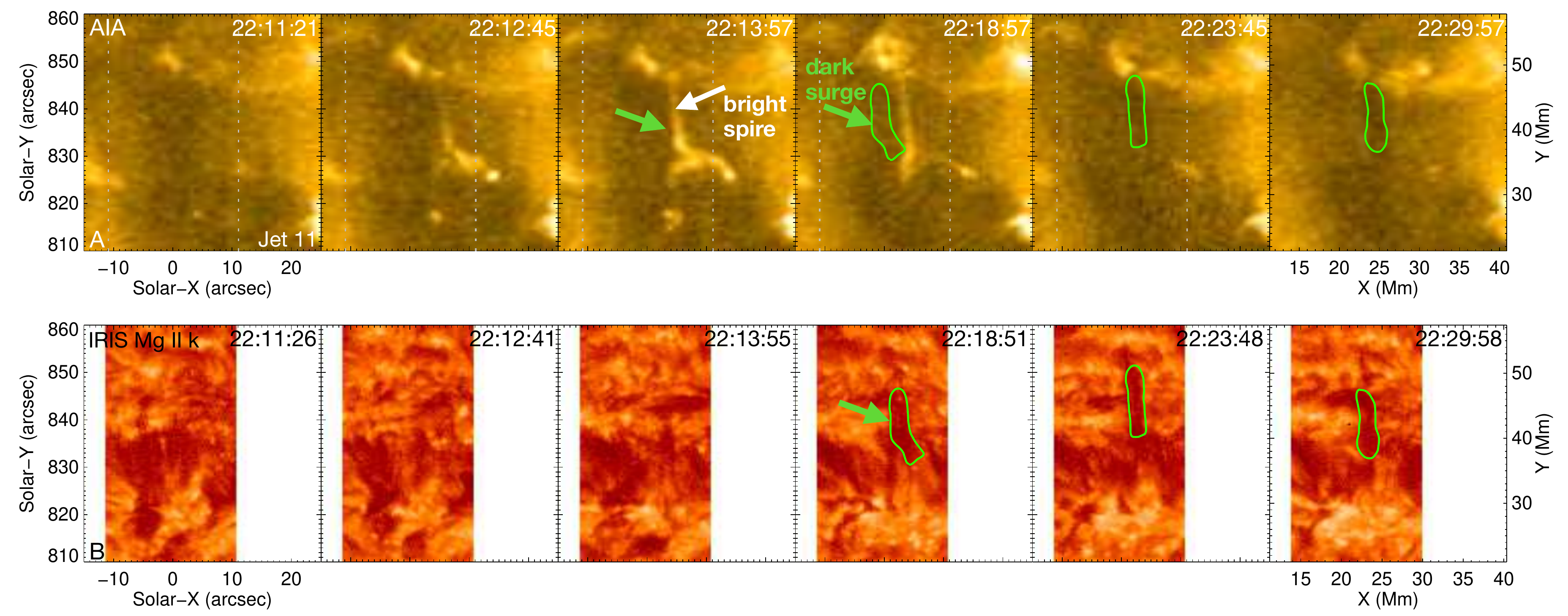}
\caption{Panel (A): Temporal evolution of Jet 11 observed in the AIA 171 \AA\ passband on August 7, 2019. The jet base exhibits a distinct inverted “Y” (or “$\lambda$”) morphology, most clearly visible at the 22:13:57 timestamp. Panel (B): coordinated IRIS observation showing the temporal evolution of raster maps, constructed by extracting intensity at 2796.2 \AA\ in the \ion{Mg}{2} k spectral window. The green contour outlines the dark surge observed in \ion{Mg}{2} k 2796.2 \AA\ and is overplotted in the AIA 171 \AA\ images in panel (A), revealing that the dark surge appears adjacent to the bright jet spire. The white vertical dashed line in the AIA panels marks the FOV of the IRIS raster maps. An animation of this figure is available online. The real-time duration of the animation is 5~s. }
\label{fig:5}
\end{figure*}
\subsection{Kinetic Energy of Jets}
We further estimated jet kinetic energies for the bright spire using the standard relation $E_k = \tfrac{1}{2}\, n_e m_p v^2 V$, with the jet volume approximated as $V = \tfrac{\pi L w^2}{4}$ where \(L\) and \(w\) are the measured jet length and width. Assuming a coronal-hole electron density of \(n_e = 2\times10^{8}\ \mathrm{cm^{-3}}\) \citep{1978ApJ...226..674F, 2023Sci...381..867C}, the resulting kinetic energies lie between \(7.30\times10^{21}\,\mathrm{erg}\) and \(9.98\times10^{22}\,\mathrm{erg}\) (mean \(:3.98\times10^{22}\,\mathrm{erg}\)). These values place our jets above the lower limit reported for picoflare jets (\(\gtrsim10^{21}\,\mathrm{erg}\); \citealt{2023Sci...381..867C}) but below the total energies quoted for jetlets (\(10^{24}-10^{25}\,\mathrm{erg}\); \citealt{2023ApJ...945...28R, 2024ApJ...963....4S}). Energetically, our jets are therefore consistent with the picoflare energy regime, while their spatial sizes and lifetimes are consistent with jetlets. We note that these kinetic energy estimates depend on the assumed density, volume, projection, and length estimation, and are therefore subject to uncertainties.
\subsection{Comparison with Flux Emergence Simulation}
We further examined the temporal evolution of jets and compared them with the flux-emergence simulation of \citet{2017ApJ...850..153N, 2018ApJ...858....8N}, introduced in section~\ref{s-NE}. Figure~\ref{fig:4} presents three representative observational cases (Jets 10, 1, and 7) in panels (A)-(C), followed by the synthetic emission in EUI 174 \AA\ filter of the simulation (panel D), temperature maps (panel E), and grayscale density maps (panel F). Panel (A) shows the temporal evolution of Jet~10 in HRI$_{\mathrm{EUV}}$ on April~5, 2024. At 21:47:56~UT, no jet-like structure is visible. By 21:50:52~UT, we start to see the appearance of a jet. At 21:51:56~UT, a JBP (red arrow) forms at the base, with a current sheet (green arrow) visible just above it, together with the bright spire (cyan arrow) and the associated upward-moving dark structure (white arrow). The jet activity in observations ceases around 22:01:00~UT. For Jet~1, observed during the north polar observation on September 14, 2021, we have shown five snapshots in panel (B). At 03:33:52~UT, the region appeared quiet, with no noticeable jet activity. By 03:35:52~UT, a JBP (red arrow) had formed at the base with a current sheet above it (green arrow), along with a developing bright spire (cyan arrow). A narrow dark structure (white arrow) appeared on the right side of the spire and became more evident at 03:36:32~UT. By 03:38:12~UT, the dark structure had faded, although the bright spire remained visible. Later, at 03:39:52~UT, other dynamic structures emerged in the vicinity, indicating ongoing activity in the region. This jet has also been studied previously by \citet{2022A&A...664A..28M} using AIA data.\\
The temporal evolution of Jet~7, observed on March 30, 2022, is shown in panel (C). At 04:48:42~UT, no significant jet activity was observed. A few minutes later, at 04:51:45~UT, a JBP (red arrow) had formed at the base with a current sheet (green arrow) above it, together with a faint loop-like structure and a developing jet spire (cyan arrow). By 04:53:48~UT, the spire became clearer, and a dark structure (white arrow) appeared on the right side. The dark structure became more distinct by 04:55:00~UT and gradually faded by 04:59:45~UT. Another jet is visible at 04:48:42~UT near coordinates (15, 5); this corresponds to Jet~4, which is also included in our analysis.
The jets in our observations follow a characteristic temporal evolution: an initially quiescent region, the appearance of a jet bright point (JBP) at the base, the formation of a current sheet above it, the development of a bright spire accompanied by dark structures, and finally a gradual fading of activity. Overall, the jets closely resemble the evolution of classical jets \citep{1994ApJ...431L..51S, 1995Natur.375...42Y} and exhibit the characteristic inverted `Y' (or `$\lambda$') morphology.\\
Panel~(D) shows the synthetic emission in the EUI 174 \AA\ passband from the Bifrost simulation, expressed in units of DN cm$^{-1}$ s$^{-1}$ pix$^{-1}$. The jet displays an inverted Y-shaped morphology and a temporal evolution closely resembling our observations. The blue contours at T\,=\,0.1\,MK trace the transition region and aid the identification of the cool ejection, while the red contours at T\,=\,2\,MK outline the hot plasma around the JBP, current sheet, and spire. The spire at T\,=\,2\,MK, however, appears dark. This is mainly due to the 2.5D nature of the simulation and the temperature response of the EUI passband, which is centered at 174 \AA\ and is most sensitive to plasma near $\sim$1 MK through \ion{Fe}{9} and \ion{Fe}{10} emission lines. So, plasma that is either significantly cooler or hotter than this peak contributes little to the 174 \AA\ intensity, and thus appears dark. Therefore, we have shown the corresponding temperature maps in panel~(E), which clearly confirm the presence of $\sim$2 MK spire plasma and the $\sim$0.1 MK cool ejection even though they appear dark in the synthetic emission. Panel~(F) shows the density maps in grayscale, in units of g cm$^{-3}$, overplotted with the same temperature contours. These maps reveal that the bright spire corresponds to low-density, high-temperature coronal plasma, whereas the adjacent dark structures on the opposite side of the JBP are associated with higher-density, lower-temperature chromospheric plasma. This close resemblance between the observed and simulated jet morphology and dynamics suggests that the studied jets are consistent with being driven by magnetic flux emergence and reconnection (\citealt{1996PASJ...48..353Y, 2012ApJ...751..152J, 2013ApJ...771...20M, 2016ApJ...822...18N, 2018ApJ...852...16Y}).\\
We emphasize that, in this study, the simulation is used primarily to provide a qualitative morphological comparison with the observed jets, rather than to establish a one-to-one correspondence. In this simulation, a relatively larger jet is produced owing to the imposed boundary conditions. Due to the finite extent of the computational domain and the limited duration of the run, the jet’s evolution is restricted by the upper boundary and the simulated time span. The upper boundary of the domain is located at $\sim$30 Mm, which allows plasma to leave the simulation domain. As a result, the full lifetime and length of the simulated jet were not captured. Nevertheless, within these constraints, we estimate the kinetic energy using the portion of the jet spire contained within the domain at a temperature of 1.5\,MK. The spire length is measured from the reconnection site to the top of the visible spire ($\sim$23 Mm) and a width of $\sim$1 Mm in the y-direction. Assuming a cylindrical geometry, we find that at the 66.67 min timestamp, when the jet reaches its maximum extent within this domain, the kinetic energy of its spire is $\sim8.15\times10^{24}$ erg. This value is comparable to the kinetic energy range derived for the jetlets \citep{2023ApJ...945...28R, 2024ApJ...963....4S} and is about two orders of magnitude higher than that of the observed jets reported here. Note that the simulation models a single, slightly larger 2.5D jet, whereas the observed jets cover smaller spatial scales with substantial event-to-event variability. Moreover, projection effects can lead to underestimated velocities and lengths in the observations. These differences in spatial scale, dimensionality, and magnetic configuration limit the extent to which a direct quantitative comparison can be made.\\
We note, however, that alternative mechanisms, such as minifilament eruptions, cannot be ruled out. The lack of sufficiently high-resolution observations prevents us from unambiguously determining whether the dark material observed in EUV images corresponds to filamentary plasma or not. Moreover, the absence of magnetic field measurements precludes us from establishing whether these jets are driven by flux cancellation or flux emergence. Some of the jets we visually inspected could be related to minifilament eruptions, as reported in earlier studies using HRI$_{\mathrm{EUV}}$ observations (\citealt{2022A&A...664A..28M, 2023ApJ...944...19L}). However, in the absence of radiative-MHD simulations for the minifilament scenario, our comparison is limited to available flux-emergence simulations, and the one-to-one correspondence between our observations and these simulations suggests that flux emergence might be responsible for driving the jets in our study. While future minifilament eruption simulations will allow direct comparisons and provide valuable complementary insights, coordinated high-resolution limb magnetograms and chromospheric imaging (H$\alpha$, \ion{Ca}{2}, \ion{Mg}{2}), combined with multi-wavelength EUV observations and Doppler diagnostics (e.g., \citealt{2020ApJ...902....8C}), will also be essential to gain critical insights into the minifilament scenario.
\subsection{Chromospheric Nature of Dark Structure}
These dark structures share key morphological and dynamical properties with previously reported chromospheric surges (\citealt{1973SoPh...32..139R, 1996ApJ...464.1016C, 2007A&A...469..331J}). With the unprecedented spatial resolution and high cadence of HRI$_{\mathrm{EUV}}$, we analyzed ten tiny coronal jets observed at the solar poles in EUV. To complement our analysis, we also utilize the coordinated IRIS and AIA observations (dataset D5), which provide diagnostics in chromospheric lines such as \ion{Mg}{2}, allowing us to study the dark structure seen in our tiny coronal jets in greater detail alongside the EUV data.\\
Figure~\ref{fig:5} shows the evolution of Jet 11 in these coordinated AIA and IRIS observations. Panel~(A) shows the evolution of Jet~11 in AIA 171~\AA. To analyze its chromospheric component, we used the \ion{Mg}{2}~{k} 2796 \AA\ spectral window from IRIS, which primarily samples chromospheric plasma at temperatures of log T\,=\,3.7\,-\,4.2\,K. Afterwards, we created intensity raster maps by extracting the intensity at 2796.2~\AA, which corresponds to the position of the k2v emission peak, from each slit position as IRIS scanned across the region. Each full map was made at a cadence of 74\,s. In the raster maps, a dark collimated structure, identified as a ‘chromospheric surge’ (highlighted by a green contour), is observed. When the same contour is overplotted on the AIA images, the surge lies immediately adjacent to the bright jet spire, on the side opposite the JBP (marked with a white arrow). The jet exhibits an inverted-Y (or ‘$\lambda$’) morphology, with a narrow spire emerging from a broader base, and the surge detected in IRIS aligns closely with the bright spire in AIA. This configuration is consistent with the standard flux-emergence scenario, as reproduced in the numerical simulations presented in this paper and also in previous studies (\citealt{1995Natur.375...42Y, 1996PASJ...48..353Y, 2012ApJ...751..152J, 2013ApJ...771...20M, 2018ApJ...852...16Y}). This one-to-one spatial correspondence suggests that the dark structure in our EUV observations might be a chromospheric surge associated with a coronal jet.\\
The identification as a surge is supported by its morphology and the gradual, collimated rise of cool and dark plasma gradually appearing and disappearing at the same location, which are characteristic features described in earlier studies of surges (\citealt{1973SoPh...32..139R, 1984SoPh...94..133S, 2007A&A...469..331J, 2012ApJ...752...70U}). Since the jets in our study are located at the limb, direct and reliable measurements of the underlying photospheric magnetic field are unfortunately not available. In the future, we hope that high-resolution magnetogram observations will enable a more detailed study of these jets and provide crucial insights into their magnetic configuration.
\section{Summary}
Picoflare jets are the smallest reconnection-driven jets observed in the solar atmosphere and have been proposed as potential contributors to coronal heating and solar wind acceleration, yet their physical origin remains elusive. In this study, we examined tiny coronal jets observed off-limb in the Sun's polar regions using high-resolution observations from the HRI$_{\mathrm{EUV}}$ on board Solar Orbiter. These jets, which we identify as picoflare-scale events, often display the simultaneous presence of a bright spire and a narrow, collimated dark structure. Our analysis shows that their spatial (length:\,7.85\,-\,29.74\,Mm; width:\,0.35\,-\,1.75\,Mm) and temporal scales (3.4\,-\,11.5 minutes) are comparable to those of jetlets, while their kinetic energies (7.30$\times10^{21}$\,-\,9.98$\times10^{22}$\,erg) fall within the picoflare regime. By combining coordinated observations from IRIS and AIA on board SDO with insights from Bifrost radiative-MHD simulations, we find a clear one-to-one morphological correspondence between the dark counterparts and cool chromospheric surges accompanying the bright spire. Our findings suggest that these small-scale, jetlet-like picoflare jets are driven by magnetic reconnection following magnetic flux emergence, resulting in coupled bright and dark plasma ejections. While coupled bright and dark plasma ejections are well documented in larger coronal jets, our results show that they also occur at the picoflare-jet scale, suggesting a unifying reconnection-driven mechanism across jet regimes. These results underscore the importance of high-resolution observations and models in resolving the fundamental components of solar activity and provide new constraints on the role of small-scale jets in coronal heating and solar wind acceleration.
\begin{acknowledgments}
Solar Orbiter is a space mission with international collaboration between ESA and NASA, operated by ESA. The EUI instrument was built by CSL, IAS, MPS, MSSL/UCL, PMOD/WRC, ROB, LCF/IO with funding from the Belgian Federal Science Policy Oﬃce (BELSPO/PRODEX PEA 4000134088); the Centre National d’Etudes Spatiales (CNES); the UK Space Agency (UKSA); the Bundesministerium für Wirtschaft und Energie (BMWi) through the Deutsches Zentrum für Luft- und Raumfahrt (DLR); and the Swiss Space Oﬃce (SSO). IRIS is a NASA small explorer mission
developed and operated by LMSAL with mission operations
executed at NASA Ames Research Center and major
contributions to downlink communications funded by ESA
and the Norwegian Space Centre. D.N.S gratefully acknowledges the support by the European Research Council through the Synergy Grant number 810218 (``The Whole Sun'', ERC-2018-SyG) and by the Research Council of Norway (RCN) through its Centres of Excellence scheme, project number 262622. J.J. acknowledges funding support from the SERB-CRG grant (CRG/2023/007464) provided by the Anusandhan National Research Foundation, India.
\end{acknowledgments}
\facilities{Solar Orbiter (EUI), IRIS,
            SDO (AIA)}
\software{IDL \citep{idl2000}} 

\bibliography{euijets}{}

@ARTICLE{2018ApJ...858....8N,
       author = {{N{\'o}brega-Siverio}, D. and {Moreno-Insertis}, F. and {Mart{\'\i}nez-Sykora}, J.},
        title = "{On the Importance of the Nonequilibrium Ionization of Si IV and O IV and the Line of Sight in Solar Surges}",
      journal = {\apj},
         year = 2018,
        month = may,
       volume = {858},
       number = {1},
          eid = {8},
        pages = {8},
          doi = {10.3847/1538-4357/aab9b9},
archivePrefix = {arXiv},
       eprint = {1803.10251},
 primaryClass = {astro-ph.SR},
       adsurl = {https://ui.adsabs.harvard.edu/abs/2018ApJ...858....8N}
}

@ARTICLE{2021A&A...656L...4B,
       author = {{Berghmans}, D. and {Auch{\`e}re}, F. and {Long}, D.~M. and {Soubri{\'e}}, E. and {Mierla}, M. and {Zhukov}, A.~N. and {Sch{\"u}hle}, U. and {Antolin}, P. and {Harra}, L. and {Parenti}, S. and {Podladchikova}, O. and {Aznar Cuadrado}, R. and {Buchlin}, {\'E}. and {Dolla}, L. and {Verbeeck}, C. and {Gissot}, S. and {Teriaca}, L. and {Haberreiter}, M. and {Katsiyannis}, A.~C. and {Rodriguez}, L. and {Kraaikamp}, E. and {Smith}, P.~J. and {Stegen}, K. and {Rochus}, P. and {Halain}, J.~P. and {Jacques}, L. and {Thompson}, W.~T. and {Inhester}, B.},
        title = "{Extreme-UV quiet Sun brightenings observed by the Solar Orbiter/EUI}",
      journal = {\aap},
         year = 2021,
        month = dec,
       volume = {656},
          eid = {L4},
        pages = {L4},
          doi = {10.1051/0004-6361/202140380},
archivePrefix = {arXiv},
       eprint = {2104.03382},
 primaryClass = {astro-ph.SR},
       adsurl = {https://ui.adsabs.harvard.edu/abs/2021A&A...656L...4B}
}

@ARTICLE{2023Sci...381..867C,
       author = {{Chitta}, L.~P. and {Zhukov}, A.~N. and {Berghmans}, D. and {Peter}, H. and {Parenti}, S. and {Mandal}, S. and {Aznar Cuadrado}, R. and {Sch{\"u}hle}, U. and {Teriaca}, L. and {Auch{\`e}re}, F. and {Barczynski}, K. and {Buchlin}, {\'E}. and {Harra}, L. and {Kraaikamp}, E. and {Long}, D.~M. and {Rodriguez}, L. and {Schwanitz}, C. and {Smith}, P.~J. and {Verbeeck}, C. and {Seaton}, D.~B.},
        title = "{Picoflare jets power the solar wind emerging from a coronal hole on the Sun}",
      journal = {Science},
         year = 2023,
        month = aug,
       volume = {381},
       number = {6660},
        pages = {867-872},
          doi = {10.1126/science.ade5801},
archivePrefix = {arXiv},
       eprint = {2308.13044},
 primaryClass = {astro-ph.SR},
       adsurl = {https://ui.adsabs.harvard.edu/abs/2023Sci...381..867C}
}

@ARTICLE{2022A&A...667A.166C,
       author = {{Chitta}, L.~P. and {Peter}, H. and {Parenti}, S. and {Berghmans}, D. and {Auch{\`e}re}, F. and {Solanki}, S.~K. and {Aznar Cuadrado}, R. and {Sch{\"u}hle}, U. and {Teriaca}, L. and {Mandal}, S. and {Barczynski}, K. and {Buchlin}, {\'E}. and {Harra}, L. and {Kraaikamp}, E. and {Long}, D.~M. and {Rodriguez}, L. and {Schwanitz}, C. and {Smith}, P.~J. and {Verbeeck}, C. and {Zhukov}, A.~N. and {Liu}, W. and {Cheung}, M.~C.~M.},
        title = "{Solar coronal heating from small-scale magnetic braids}",
      journal = {\aap},
         year = 2022,
        month = nov,
       volume = {667},
          eid = {A166},
        pages = {A166},
          doi = {10.1051/0004-6361/202244170},
archivePrefix = {arXiv},
       eprint = {2209.12203},
 primaryClass = {astro-ph.SR},
       adsurl = {https://ui.adsabs.harvard.edu/abs/2022A&A...667A.166C}
}

@ARTICLE{2022A&A...664A..28M,
       author = {{Mandal}, Sudip and {Chitta}, Lakshmi Pradeep and {Peter}, Hardi and {Solanki}, Sami K. and {Cuadrado}, Regina Aznar and {Teriaca}, Luca and {Sch{\"u}hle}, Udo and {Berghmans}, David and {Auch{\`e}re}, Fr{\'e}d{\'e}ric},
        title = "{A highly dynamic small-scale jet in a polar coronal hole}",
      journal = {\aap},
         year = 2022,
        month = aug,
       volume = {664},
          eid = {A28},
        pages = {A28},
          doi = {10.1051/0004-6361/202243765},
archivePrefix = {arXiv},
       eprint = {2206.02236},
 primaryClass = {astro-ph.SR},
       adsurl = {https://ui.adsabs.harvard.edu/abs/2022A&A...664A..28M}
}

@ARTICLE{1996ApJ...464.1016C,
       author = {{Canfield}, Richard C. and {Reardon}, Kevin P. and {Leka}, K.~D. and {Shibata}, K. and {Yokoyama}, T. and {Shimojo}, M.},
        title = "{H alpha Surges and X-Ray Jets in AR 7260}",
      journal = {\apj},
         year = 1996,
        month = jun,
       volume = {464},
        pages = {1016},
          doi = {10.1086/177389},
       adsurl = {https://ui.adsabs.harvard.edu/abs/1996ApJ...464.1016C}
}

@ARTICLE{1992PASJ...44L.173S,
       author = {{Shibata}, Kazunari and {Ishido}, Yoshinori and {Acton}, Loren W. and {Strong}, Keith T. and {Hirayama}, Tadashi and {Uchida}, Yutaka and {McAllister}, Alan H. and {Matsumoto}, Ryoji and {Tsuneta}, Saku and {Shimizu}, Toshifumi and {Hara}, Hirohisa and {Sakurai}, Takashi and {Ichimoto}, Kiyoshi and {Nishino}, Yohei and {Ogawara}, Yoshiaki},
        title = "{Observations of X-Ray Jets with the YOHKOH Soft X-Ray Telescope}",
      journal = {\pasj},
         year = 1992,
        month = oct,
       volume = {44},
        pages = {L173-L179},
       adsurl = {https://ui.adsabs.harvard.edu/abs/1992PASJ...44L.173S}
}

@ARTICLE{2016ApJ...822...18N,
       author = {{N{\'o}brega-Siverio}, D. and {Moreno-Insertis}, F. and {Mart{\'\i}nez-Sykora}, J.},
        title = "{The Cool Surge Following Flux Emergence in a Radiation-MHD Experiment}",
      journal = {\apj},
         year = 2016,
        month = may,
       volume = {822},
       number = {1},
          eid = {18},
        pages = {18},
          doi = {10.3847/0004-637X/822/1/18},
archivePrefix = {arXiv},
       eprint = {1601.04074},
 primaryClass = {astro-ph.SR},
       adsurl = {https://ui.adsabs.harvard.edu/abs/2016ApJ...822...18N}
}

@ARTICLE{1997Natur.386..811I,
       author = {{Innes}, D.~E. and {Inhester}, B. and {Axford}, W.~I. and {Wilhelm}, K.},
        title = "{Bi-directional plasma jets produced by magnetic reconnection on the Sun}",
      journal = {\nat},
         year = 1997,
        month = apr,
       volume = {386},
       number = {6627},
        pages = {811-813},
          doi = {10.1038/386811a0},
       adsurl = {https://ui.adsabs.harvard.edu/abs/1997Natur.386..811I}
}

@ARTICLE{2021RSPSA.47700217S,
       author = {{Shen}, Yuandeng},
        title = "{Observation and modelling of solar jets}",
      journal = {Proceedings of the Royal Society of London Series A},
         year = 2021,
        month = feb,
       volume = {477},
       number = {2246},
        pages = {217},
          doi = {10.1098/rspa.2020.0217},
archivePrefix = {arXiv},
       eprint = {2101.04846},
 primaryClass = {astro-ph.SR},
       adsurl = {https://ui.adsabs.harvard.edu/abs/2021RSPSA.47700217S}
}

@ARTICLE{1996PASJ...48..123S,
       author = {{Shimojo}, Masumi and {Hashimoto}, Shizuyo and {Shibata}, Kazunari and {Hirayama}, Tadashi and {Hudson}, Hugh S. and {Acton}, Loren W.},
        title = "{Statistical Study of Solar X-Ray Jets Observed with the YOHKOH Soft X-Ray Telescope}",
      journal = {\pasj},
         year = 1996,
        month = feb,
       volume = {48},
        pages = {123-136},
          doi = {10.1093/pasj/48.1.123},
       adsurl = {https://ui.adsabs.harvard.edu/abs/1996PASJ...48..123S}
}

@ARTICLE{2015Natur.523..437S,
       author = {{Sterling}, Alphonse C. and {Moore}, Ronald L. and {Falconer}, David A. and {Adams}, Mitzi},
        title = "{Small-scale filament eruptions as the driver of X-ray jets in solar coronal holes}",
      journal = {\nat},
         year = 2015,
        month = jul,
       volume = {523},
       number = {7561},
        pages = {437-440},
          doi = {10.1038/nature14556},
archivePrefix = {arXiv},
       eprint = {1705.03373},
 primaryClass = {astro-ph.SR},
       adsurl = {https://ui.adsabs.harvard.edu/abs/2015Natur.523..437S}
}

@ARTICLE{2016ApJ...828L...9S,
       author = {{Sterling}, Alphonse C. and {Moore}, Ronald L.},
        title = "{A Microfilament-eruption Mechanism for Solar Spicules}",
      journal = {\apjl},
         year = 2016,
        month = sep,
       volume = {828},
       number = {1},
          eid = {L9},
        pages = {L9},
          doi = {10.3847/2041-8205/828/1/L9},
archivePrefix = {arXiv},
       eprint = {1612.00430},
 primaryClass = {astro-ph.SR},
       adsurl = {https://ui.adsabs.harvard.edu/abs/2016ApJ...828L...9S}
}

@ARTICLE{2009SoPh..259...87N,
       author = {{Nistic{\`o}}, G. and {Bothmer}, V. and {Patsourakos}, S. and {Zimbardo}, G.},
        title = "{Characteristics of EUV Coronal Jets Observed with STEREO/SECCHI}",
      journal = {\solphys},
         year = 2009,
        month = oct,
       volume = {259},
       number = {1-2},
          eid = {87},
        pages = {87},
          doi = {10.1007/s11207-009-9424-8},
archivePrefix = {arXiv},
       eprint = {0906.4407},
 primaryClass = {astro-ph.SR},
       adsurl = {https://ui.adsabs.harvard.edu/abs/2009SoPh..259...87N}
}

@ARTICLE{2025A&A...702A.188N,
       author = {{N{\'o}brega-Siverio}, Daniel and {Joshi}, Reetika and {Sola-Viladesau}, Eva and {Berghmans}, David and {Lim}, Daye},
        title = "{Thin coronal jets and plasmoid-mediated reconnection: Insights from Solar Orbiter observations and Bifrost simulations}",
      journal = {\aap},
         year = 2025,
        month = oct,
       volume = {702},
          eid = {A188},
        pages = {A188},
          doi = {10.1051/0004-6361/202555357},
archivePrefix = {arXiv},
       eprint = {2506.03092},
 primaryClass = {astro-ph.SR},
       adsurl = {https://ui.adsabs.harvard.edu/abs/2025A&A...702A.188N}
}

@ARTICLE{2023ApJ...945...28R,
       author = {{Raouafi}, Nour E. and {Stenborg}, G. and {Seaton}, D.~B. and {Wang}, H. and {Wang}, J. and {DeForest}, C.~E. and {Bale}, S.~D. and {Drake}, J.~F. and {Uritsky}, V.~M. and {Karpen}, J.~T. and {DeVore}, C.~R. and {Sterling}, A.~C. and {Horbury}, T.~S. and {Harra}, L.~K. and {Bourouaine}, S. and {Kasper}, J.~C. and {Kumar}, P. and {Phan}, T.~D. and {Velli}, M.},
        title = "{Magnetic Reconnection as the Driver of the Solar Wind}",
      journal = {\apj},
         year = 2023,
        month = mar,
       volume = {945},
       number = {1},
          eid = {28},
        pages = {28},
          doi = {10.3847/1538-4357/acaf6c},
archivePrefix = {arXiv},
       eprint = {2301.00903},
 primaryClass = {astro-ph.SR},
       adsurl = {https://ui.adsabs.harvard.edu/abs/2023ApJ...945...28R}
}

@ARTICLE{2016ApJ...821..100S,
       author = {{Sterling}, Alphonse C. and {Moore}, Ronald L. and {Falconer}, David A. and {Panesar}, Navdeep K. and {Akiyama}, Sachiko and {Yashiro}, Seiji and {Gopalswamy}, Nat},
        title = "{Minifilament Eruptions that Drive Coronal Jets in a Solar Active Region}",
      journal = {\apj},
         year = 2016,
        month = apr,
       volume = {821},
       number = {2},
          eid = {100},
        pages = {100},
          doi = {10.3847/0004-637X/821/2/100},
       adsurl = {https://ui.adsabs.harvard.edu/abs/2016ApJ...821..100S}
}

@ARTICLE{2011ApJ...738L..20H,
       author = {{Hong}, Junchao and {Jiang}, Yunchun and {Zheng}, Ruisheng and {Yang}, Jiayan and {Bi}, Yi and {Yang}, Bo},
        title = "{A Micro Coronal Mass Ejection Associated Blowout Extreme-ultraviolet Jet}",
      journal = {\apjl},
         year = 2011,
        month = sep,
       volume = {738},
       number = {2},
          eid = {L20},
        pages = {L20},
          doi = {10.1088/2041-8205/738/2/L20},
       adsurl = {https://ui.adsabs.harvard.edu/abs/2011ApJ...738L..20H}
}

@ARTICLE{2019Sci...366..890S,
       author = {{Samanta}, Tanmoy and {Tian}, Hui and {Yurchyshyn}, Vasyl and {Peter}, Hardi and {Cao}, Wenda and {Sterling}, Alphonse and {Erd{\'e}lyi}, Robertus and {Ahn}, Kwangsu and {Feng}, Song and {Utz}, Dominik and {Banerjee}, Dipankar and {Chen}, Yajie},
        title = "{Generation of solar spicules and subsequent atmospheric heating}",
      journal = {Science},
         year = 2019,
        month = nov,
       volume = {366},
       number = {6467},
        pages = {890-894},
          doi = {10.1126/science.aaw2796},
archivePrefix = {arXiv},
       eprint = {2006.02571},
 primaryClass = {astro-ph.SR},
       adsurl = {https://ui.adsabs.harvard.edu/abs/2019Sci...366..890S}
}

@ARTICLE{2025ApJ...988L..65B,
       author = {{Bura}, Annu and {Shrivastav}, Arpit Kumar and {Patel}, Ritesh and {Samanta}, Tanmoy and {Nayak}, Sushree S. and {Ghosh}, Ananya and {Sow Mondal}, Shanwlee and {Pant}, Vaibhav and {Seaton}, Daniel B.},
        title = "{Dynamics of Reconnection Nanojets in Eruptive and Confined Solar Flares}",
      journal = {\apjl},
         year = 2025,
        month = aug,
       volume = {988},
       number = {2},
          eid = {L65},
        pages = {L65},
          doi = {10.3847/2041-8213/adef0f},
archivePrefix = {arXiv},
       eprint = {2507.04639},
 primaryClass = {astro-ph.SR},
       adsurl = {https://ui.adsabs.harvard.edu/abs/2025ApJ...988L..65B}
}

@ARTICLE{1973SoPh...32..139R,
       author = {{Roy}, J. -Ren{\'e}},
        title = "{The Dynamics of Solar Surges}",
      journal = {\solphys},
         year = 1973,
        month = sep,
       volume = {32},
       number = {1},
        pages = {139-151},
          doi = {10.1007/BF00152734},
       adsurl = {https://ui.adsabs.harvard.edu/abs/1973SoPh...32..139R}
}

@ARTICLE{1984SoPh...94..133S,
       author = {{Schmieder}, B. and {Mein}, P. and {Martres}, M.~J. and {Tandberg-Hanssen}, E.},
        title = "{Dynamic evolution of recurrent mass ejections observed in H{\ensuremath{\alpha}} and C iv lines}",
      journal = {\solphys},
         year = 1984,
        month = aug,
       volume = {94},
       number = {1},
        pages = {133-150},
          doi = {10.1007/BF00154814},
       adsurl = {https://ui.adsabs.harvard.edu/abs/1984SoPh...94..133S}
}

@ARTICLE{2007A&A...469..331J,
       author = {{Jiang}, Y.~C. and {Chen}, H.~D. and {Li}, K.~J. and {Shen}, Y.~D. and {Yang}, L.~H.},
        title = "{The H{\ensuremath{\alpha}} surges and EUV jets from magnetic flux emergences and cancellations}",
      journal = {\aap},
         year = 2007,
        month = jul,
       volume = {469},
       number = {1},
        pages = {331-337},
          doi = {10.1051/0004-6361:20053954},
       adsurl = {https://ui.adsabs.harvard.edu/abs/2007A&A...469..331J}
}

@ARTICLE{2014ApJ...794..140V,
       author = {{Vargas Dom{\'\i}nguez}, Santiago and {Kosovichev}, Alexander and {Yurchyshyn}, Vasyl},
        title = "{Multi-wavelength High-resolution Observations of a Small-scale Emerging Magnetic Flux Event and the Chromospheric and Coronal Response}",
      journal = {\apj},
         year = 2014,
        month = oct,
       volume = {794},
       number = {2},
          eid = {140},
        pages = {140},
          doi = {10.1088/0004-637X/794/2/140},
archivePrefix = {arXiv},
       eprint = {1405.3550},
 primaryClass = {astro-ph.SR},
       adsurl = {https://ui.adsabs.harvard.edu/abs/2014ApJ...794..140V}
}

@ARTICLE{2012ApJ...752...70U,
       author = {{Uddin}, Wahab and {Schmieder}, B. and {Chandra}, R. and {Srivastava}, Abhishek K. and {Kumar}, Pankaj and {Bisht}, S.},
        title = "{Observations of Multiple Surges Associated with Magnetic Activities in AR 10484 on 2003 October 25}",
      journal = {\apj},
         year = 2012,
        month = jun,
       volume = {752},
       number = {1},
          eid = {70},
        pages = {70},
          doi = {10.1088/0004-637X/752/1/70},
archivePrefix = {arXiv},
       eprint = {1204.2053},
 primaryClass = {astro-ph.SR},
       adsurl = {https://ui.adsabs.harvard.edu/abs/2012ApJ...752...70U}
}

@ARTICLE{2020ApJ...896L..18S,
       author = {{Sterling}, Alphonse C. and {Moore}, Ronald L.},
        title = "{Coronal-jet-producing Minifilament Eruptions as a Possible Source of Parker Solar Probe Switchbacks}",
      journal = {\apjl},
         year = 2020,
        month = jun,
       volume = {896},
       number = {2},
          eid = {L18},
        pages = {L18},
          doi = {10.3847/2041-8213/ab96be},
archivePrefix = {arXiv},
       eprint = {2006.04990},
 primaryClass = {astro-ph.SR},
       adsurl = {https://ui.adsabs.harvard.edu/abs/2020ApJ...896L..18S}
}

@ARTICLE{2012SoPh..275...17L,
       author = {{Lemen}, James R. and {Title}, Alan M. and {Akin}, David J. and {Boerner}, Paul F. and {Chou}, Catherine and {Drake}, Jerry F. and {Duncan}, Dexter W. and {Edwards}, Christopher G. and {Friedlaender}, Frank M. and {Heyman}, Gary F. and {Hurlburt}, Neal E. and {Katz}, Noah L. and {Kushner}, Gary D. and {Levay}, Michael and {Lindgren}, Russell W. and {Mathur}, Dnyanesh P. and {McFeaters}, Edward L. and {Mitchell}, Sarah and {Rehse}, Roger A. and {Schrijver}, Carolus J. and {Springer}, Larry A. and {Stern}, Robert A. and {Tarbell}, Theodore D. and {Wuelser}, Jean-Pierre and {Wolfson}, C. Jacob and {Yanari}, Carl and {Bookbinder}, Jay A. and {Cheimets}, Peter N. and {Caldwell}, David and {Deluca}, Edward E. and {Gates}, Richard and {Golub}, Leon and {Park}, Sang and {Podgorski}, William A. and {Bush}, Rock I. and {Scherrer}, Philip H. and {Gummin}, Mark A. and {Smith}, Peter and {Auker}, Gary and {Jerram}, Paul and {Pool}, Peter and {Soufli}, Regina and {Windt}, David L. and {Beardsley}, Sarah and {Clapp}, Matthew and {Lang}, James and {Waltham}, Nicholas},
        title = "{The Atmospheric Imaging Assembly (AIA) on the Solar Dynamics Observatory (SDO)}",
      journal = {\solphys},
         year = 2012,
        month = jan,
       volume = {275},
       number = {1-2},
        pages = {17-40},
          doi = {10.1007/s11207-011-9776-8},
       adsurl = {https://ui.adsabs.harvard.edu/abs/2012SoPh..275...17L}
}

@ARTICLE{2012ApJ...745..164S,
       author = {{Shen}, Yuandeng and {Liu}, Yu and {Su}, Jiangtao and {Deng}, Yuanyong},
        title = "{On a Coronal Blowout Jet: The First Observation of a Simultaneously Produced Bubble-like CME and a Jet-like CME in a Solar Event}",
      journal = {\apj},
         year = 2012,
        month = feb,
       volume = {745},
       number = {2},
          eid = {164},
        pages = {164},
          doi = {10.1088/0004-637X/745/2/164},
archivePrefix = {arXiv},
       eprint = {1110.5243},
 primaryClass = {astro-ph.SR},
       adsurl = {https://ui.adsabs.harvard.edu/abs/2012ApJ...745..164S}
}

@ARTICLE{2014ApJ...783...11A,
       author = {{Adams}, Mitzi and {Sterling}, Alphonse C. and {Moore}, Ronald L. and {Gary}, G. Allen},
        title = "{A Small-scale Eruption Leading to a Blowout Macrospicule Jet in an On-disk Coronal Hole}",
      journal = {\apj},
         year = 2014,
        month = mar,
       volume = {783},
       number = {1},
          eid = {11},
        pages = {11},
          doi = {10.1088/0004-637X/783/1/11},
       adsurl = {https://ui.adsabs.harvard.edu/abs/2014ApJ...783...11A}
}

@ARTICLE{2009SoPh..255...79C,
       author = {{Chen}, Huadong and {Jiang}, Yunchun and {Ma}, Suli},
        title = "{An EUV Jet and H{\ensuremath{\alpha}} Filament Eruption Associated with Flux Cancelation in a Decaying Active Region}",
      journal = {\solphys},
         year = 2009,
        month = mar,
       volume = {255},
       number = {1},
        pages = {79-90},
          doi = {10.1007/s11207-008-9298-1},
       adsurl = {https://ui.adsabs.harvard.edu/abs/2009SoPh..255...79C}
}

@ARTICLE{2017A&A...606A...4M,
       author = {{Mulay}, Sargam M. and {Del Zanna}, Giulio and {Mason}, Helen},
        title = "{Cool and hot emission in a recurring active region jet}",
      journal = {\aap},
         year = 2017,
        month = sep,
       volume = {606},
          eid = {A4},
        pages = {A4},
          doi = {10.1051/0004-6361/201730429},
       adsurl = {https://ui.adsabs.harvard.edu/abs/2017A&A...606A...4M}
}

@ARTICLE{1999ApJ...513L..75C,
       author = {{Chae}, Jongchul and {Qiu}, Jiong and {Wang}, Haimin and {Goode}, Philip R.},
        title = "{Extreme-Ultraviolet Jets and H{\ensuremath{\alpha}} Surges in Solar Microflares}",
      journal = {\apjl},
         year = 1999,
        month = mar,
       volume = {513},
       number = {1},
        pages = {L75-L78},
          doi = {10.1086/311910},
       adsurl = {https://ui.adsabs.harvard.edu/abs/1999ApJ...513L..75C}
}

@ARTICLE{2018ApJ...852...16Y,
       author = {{Yang}, Liping and {Peter}, Hardi and {He}, Jiansen and {Tu}, Chuanyi and {Wang}, Linghua and {Zhang}, Lei and {Yan}, Limei},
        title = "{Formation of Cool and Warm Jets by Magnetic Flux Emerging from the Solar Chromosphere to Transition Region}",
      journal = {\apj},
         year = 2018,
        month = jan,
       volume = {852},
       number = {1},
          eid = {16},
        pages = {16},
          doi = {10.3847/1538-4357/aa9996},
       adsurl = {https://ui.adsabs.harvard.edu/abs/2018ApJ...852...16Y}
}

@ARTICLE{2020ApJ...889..187S,
       author = {{Sterling}, Alphonse C. and {Moore}, Ronald L. and {Panesar}, Navdeep K. and {Reardon}, Kevin P. and {Molnar}, Momchil and {Rachmeler}, Laurel A. and {Savage}, Sabrina L. and {Winebarger}, Amy R.},
        title = "{Hi-C 2.1 Observations of Small-scale Miniature-filament-eruption-like Cool Ejections in an Active Region Plage}",
      journal = {\apj},
         year = 2020,
        month = feb,
       volume = {889},
       number = {2},
          eid = {187},
        pages = {187},
          doi = {10.3847/1538-4357/ab5dcc},
archivePrefix = {arXiv},
       eprint = {1912.02319},
 primaryClass = {astro-ph.SR},
       adsurl = {https://ui.adsabs.harvard.edu/abs/2020ApJ...889..187S}
}

@ARTICLE{2017ApJ...851...67S,
       author = {{Shen}, Yuandeng and {Liu}, Ying D. and {Su}, Jiangtao and {Qu}, Zhining and {Tian}, Zhanjun},
        title = "{On a Solar Blowout Jet: Driving Mechanism and the Formation of Cool and Hot Components}",
      journal = {\apj},
         year = 2017,
        month = dec,
       volume = {851},
       number = {1},
          eid = {67},
        pages = {67},
          doi = {10.3847/1538-4357/aa9a48},
archivePrefix = {arXiv},
       eprint = {1711.02270},
 primaryClass = {astro-ph.SR},
       adsurl = {https://ui.adsabs.harvard.edu/abs/2017ApJ...851...67S}
}

@ARTICLE{2013ApJ...769..134M,
       author = {{Moore}, Ronald L. and {Sterling}, Alphonse C. and {Falconer}, David A. and {Robe}, Dominic},
        title = "{The Cool Component and the Dichotomy, Lateral Expansion, and Axial Rotation of Solar X-Ray Jets}",
      journal = {\apj},
         year = 2013,
        month = jun,
       volume = {769},
       number = {2},
          eid = {134},
        pages = {134},
          doi = {10.1088/0004-637X/769/2/134},
       adsurl = {https://ui.adsabs.harvard.edu/abs/2013ApJ...769..134M}
}

@ARTICLE{2012ApJ...751..152J,
       author = {{Jiang}, R. -L. and {Fang}, C. and {Chen}, P. -F.},
        title = "{Numerical Simulation of Solar Microflares in a Canopy-type Magnetic Configuration}",
      journal = {\apj},
         year = 2012,
        month = jun,
       volume = {751},
       number = {2},
          eid = {152},
        pages = {152},
          doi = {10.1088/0004-637X/751/2/152},
archivePrefix = {arXiv},
       eprint = {1204.5847},
 primaryClass = {astro-ph.SR},
       adsurl = {https://ui.adsabs.harvard.edu/abs/2012ApJ...751..152J}
}

@ARTICLE{2013ApJ...771...20M,
       author = {{Moreno-Insertis}, F. and {Galsgaard}, K.},
        title = "{Plasma Jets and Eruptions in Solar Coronal Holes: A Three-dimensional Flux Emergence Experiment}",
      journal = {\apj},
         year = 2013,
        month = jul,
       volume = {771},
       number = {1},
          eid = {20},
        pages = {20},
          doi = {10.1088/0004-637X/771/1/20},
archivePrefix = {arXiv},
       eprint = {1305.2201},
 primaryClass = {astro-ph.SR},
       adsurl = {https://ui.adsabs.harvard.edu/abs/2013ApJ...771...20M}
}

@ARTICLE{1996PASJ...48..353Y,
       author = {{Yokoyama}, Takaaki and {Shibata}, Kazunari},
        title = "{Numerical Simulation of Solar Coronal X-Ray Jets Based on the Magnetic Reconnection Model}",
      journal = {\pasj},
         year = 1996,
        month = apr,
       volume = {48},
        pages = {353-376},
          doi = {10.1093/pasj/48.2.353},
       adsurl = {https://ui.adsabs.harvard.edu/abs/1996PASJ...48..353Y}
}

@ARTICLE{2020A&A...642A...8R,
       author = {{Rochus}, P. and {Auch{\`e}re}, F. and {Berghmans}, D. and {Harra}, L. and {Schmutz}, W. and {Sch{\"u}hle}, U. and {Addison}, P. and {Appourchaux}, T. and {Aznar Cuadrado}, R. and {Baker}, D. and {Barbay}, J. and {Bates}, D. and {BenMoussa}, A. and {Bergmann}, M. and {Beurthe}, C. and {Borgo}, B. and {Bonte}, K. and {Bouzit}, M. and {Bradley}, L. and {B{\"u}chel}, V. and {Buchlin}, E. and {B{\"u}chner}, J. and {Cab{\'e}}, F. and {Cadiergues}, L. and {Chaigneau}, M. and {Chares}, B. and {Choque Cortez}, C. and {Coker}, P. and {Condamin}, M. and {Coumar}, S. and {Curdt}, W. and {Cutler}, J. and {Davies}, D. and {Davison}, G. and {Defise}, J. -M. and {Del Zanna}, G. and {Delmotte}, F. and {Delouille}, V. and {Dolla}, L. and {Dumesnil}, C. and {D{\"u}rig}, F. and {Enge}, R. and {Fran{\c{c}}ois}, S. and {Fourmond}, J. -J. and {Gillis}, J. -M. and {Giordanengo}, B. and {Gissot}, S. and {Green}, L.~M. and {Guerreiro}, N. and {Guilbaud}, A. and {Gyo}, M. and {Haberreiter}, M. and {Hafiz}, A. and {Hailey}, M. and {Halain}, J. -P. and {Hansotte}, J. and {Hecquet}, C. and {Heerlein}, K. and {Hellin}, M. -L. and {Hemsley}, S. and {Hermans}, A. and {Hervier}, V. and {Hochedez}, J. -F. and {Houbrechts}, Y. and {Ihsan}, K. and {Jacques}, L. and {J{\'e}r{\^o}me}, A. and {Jones}, J. and {Kahle}, M. and {Kennedy}, T. and {Klaproth}, M. and {Kolleck}, M. and {Koller}, S. and {Kotsialos}, E. and {Kraaikamp}, E. and {Langer}, P. and {Lawrenson}, A. and {Le Clech'}, J. -C. and {Lenaerts}, C. and {Liebecq}, S. and {Linder}, D. and {Long}, D.~M. and {Mampaey}, B. and {Markiewicz-Innes}, D. and {Marquet}, B. and {Marsch}, E. and {Matthews}, S. and {Mazy}, E. and {Mazzoli}, A. and {Meining}, S. and {Meltchakov}, E. and {Mercier}, R. and {Meyer}, S. and {Monecke}, M. and {Monfort}, F. and {Morinaud}, G. and {Moron}, F. and {Mountney}, L. and {M{\"u}ller}, R. and {Nicula}, B. and {Parenti}, S. and {Peter}, H. and {Pfiffner}, D. and {Philippon}, A. and {Phillips}, I. and {Plesseria}, J. -Y. and {Pylyser}, E. and {Rabecki}, F. and {Ravet-Krill}, M. -F. and {Rebellato}, J. and {Renotte}, E. and {Rodriguez}, L. and {Roose}, S. and {Rosin}, J. and {Rossi}, L. and {Roth}, P. and {Rouesnel}, F. and {Roulliay}, M. and {Rousseau}, A. and {Ruane}, K. and {Scanlan}, J. and {Schlatter}, P. and {Seaton}, D.~B. and {Silliman}, K. and {Smit}, S. and {Smith}, P.~J. and {Solanki}, S.~K. and {Spescha}, M. and {Spencer}, A. and {Stegen}, K. and {Stockman}, Y. and {Szwec}, N. and {Tamiatto}, C. and {Tandy}, J. and {Teriaca}, L. and {Theobald}, C. and {Tychon}, I. and {van Driel-Gesztelyi}, L. and {Verbeeck}, C. and {Vial}, J. -C. and {Werner}, S. and {West}, M.~J. and {Westwood}, D. and {Wiegelmann}, T. and {Willis}, G. and {Winter}, B. and {Zerr}, A. and {Zhang}, X. and {Zhukov}, A.~N.},
        title = "{The Solar Orbiter EUI instrument: The Extreme Ultraviolet Imager}",
      journal = {\aap},
         year = 2020,
        month = oct,
       volume = {642},
          eid = {A8},
        pages = {A8},
          doi = {10.1051/0004-6361/201936663},
       adsurl = {https://ui.adsabs.harvard.edu/abs/2020A&A...642A...8R}
}

@ARTICLE{2020A&A...642A...1M,
       author = {{M{\"u}ller}, D. and {St. Cyr}, O.~C. and {Zouganelis}, I. and {Gilbert}, H.~R. and {Marsden}, R. and {Nieves-Chinchilla}, T. and {Antonucci}, E. and {Auch{\`e}re}, F. and {Berghmans}, D. and {Horbury}, T.~S. and {Howard}, R.~A. and {Krucker}, S. and {Maksimovic}, M. and {Owen}, C.~J. and {Rochus}, P. and {Rodriguez-Pacheco}, J. and {Romoli}, M. and {Solanki}, S.~K. and {Bruno}, R. and {Carlsson}, M. and {Fludra}, A. and {Harra}, L. and {Hassler}, D.~M. and {Livi}, S. and {Louarn}, P. and {Peter}, H. and {Sch{\"u}hle}, U. and {Teriaca}, L. and {del Toro Iniesta}, J.~C. and {Wimmer-Schweingruber}, R.~F. and {Marsch}, E. and {Velli}, M. and {De Groof}, A. and {Walsh}, A. and {Williams}, D.},
        title = "{The Solar Orbiter mission. Science overview}",
      journal = {\aap},
         year = 2020,
        month = oct,
       volume = {642},
          eid = {A1},
        pages = {A1},
          doi = {10.1051/0004-6361/202038467},
archivePrefix = {arXiv},
       eprint = {2009.00861},
 primaryClass = {astro-ph.SR},
       adsurl = {https://ui.adsabs.harvard.edu/abs/2020A&A...642A...1M}
}

@ARTICLE{2021ApJ...921L..20P,
       author = {{Panesar}, Navdeep K. and {Tiwari}, Sanjiv K. and {Berghmans}, David and {Cheung}, Mark C.~M. and {M{\"u}ller}, Daniel and {Auchere}, Frederic and {Zhukov}, Andrei},
        title = "{The Magnetic Origin of Solar Campfires}",
      journal = {\apjl},
         year = 2021,
        month = nov,
       volume = {921},
       number = {1},
          eid = {L20},
        pages = {L20},
          doi = {10.3847/2041-8213/ac3007},
archivePrefix = {arXiv},
       eprint = {2110.06846},
 primaryClass = {astro-ph.SR},
       adsurl = {https://ui.adsabs.harvard.edu/abs/2021ApJ...921L..20P}
}

@ARTICLE{2022A&A...660A.143K,
       author = {{Kahil}, F. and {Hirzberger}, J. and {Solanki}, S.~K. and {Chitta}, L.~P. and {Peter}, H. and {Auch{\`e}re}, F. and {Sinjan}, J. and {Orozco Su{\'a}rez}, D. and {Albert}, K. and {Albelo Jorge}, N. and {Appourchaux}, T. and {Alvarez-Herrero}, A. and {Blanco Rodr{\'\i}guez}, J. and {Gandorfer}, A. and {Germerott}, D. and {Guerrero}, L. and {Guti{\'e}rrez M{\'a}rquez}, P. and {Kolleck}, M. and {del Toro Iniesta}, J.~C. and {Volkmer}, R. and {Woch}, J. and {Fiethe}, B. and {G{\'o}mez Cama}, J.~M. and {P{\'e}rez-Grande}, I. and {Sanchis Kilders}, E. and {Balaguer Jim{\'e}nez}, M. and {Bellot Rubio}, L.~R. and {Calchetti}, D. and {Carmona}, M. and {Deutsch}, W. and {Fern{\'a}ndez-Rico}, G. and {Fern{\'a}ndez-Medina}, A. and {Garc{\'\i}a Parejo}, P. and {Gasent-Blesa}, J.~L. and {Gizon}, L. and {Grauf}, B. and {Heerlein}, K. and {Lagg}, A. and {Lange}, T. and {L{\'o}pez Jim{\'e}nez}, A. and {Maue}, T. and {Meller}, R. and {Michalik}, H. and {Moreno Vacas}, A. and {M{\"u}ller}, R. and {Nakai}, E. and {Schmidt}, W. and {Schou}, J. and {Sch{\"u}hle}, U. and {Staub}, J. and {Strecker}, H. and {Torralbo}, I. and {Valori}, G. and {Aznar Cuadrado}, R. and {Teriaca}, L. and {Berghmans}, D. and {Verbeeck}, C. and {Kraaikamp}, E. and {Gissot}, S.},
        title = "{The magnetic drivers of campfires seen by the Polarimetric and Helioseismic Imager (PHI) on Solar Orbiter}",
      journal = {\aap},
         year = 2022,
        month = apr,
       volume = {660},
          eid = {A143},
        pages = {A143},
          doi = {10.1051/0004-6361/202142873},
archivePrefix = {arXiv},
       eprint = {2202.13859},
 primaryClass = {astro-ph.SR},
       adsurl = {https://ui.adsabs.harvard.edu/abs/2022A&A...660A.143K}
}

@ARTICLE{2025A&A...699A.138N,
       author = {{Narang}, Nancy and {Verbeeck}, Cis and {Mierla}, Marilena and {Berghmans}, David and {Auch{\`e}re}, Fr{\'e}d{\'e}ric and {Shestov}, Sergei and {Delouille}, V{\'e}ronique and {Chitta}, Lakshmi Pradeep and {Priest}, Eric and {Lim}, Daye and {Dolla}, Laurent R. and {Kraaikamp}, Emil},
        title = "{Extreme-ultraviolet transient brightenings in the quiet-Sun corona: Closest perihelion observations with Solar Orbiter/EUI}",
      journal = {\aap},
         year = 2025,
        month = jul,
       volume = {699},
          eid = {A138},
        pages = {A138},
          doi = {10.1051/0004-6361/202554650},
archivePrefix = {arXiv},
       eprint = {2505.03656},
 primaryClass = {astro-ph.SR},
       adsurl = {https://ui.adsabs.harvard.edu/abs/2025A&A...699A.138N}
}

@ARTICLE{2025ApJ...985L..12G,
       author = {{Gao}, Yuhang and {Tian}, Hui and {Berghmans}, David and {Duan}, Yadan and {Van Doorsselaere}, Tom and {Chen}, Hechao and {Kraaikamp}, Emil},
        title = "{Reconnection Nanojets in an Erupting Solar Filament with Unprecedented High Speeds}",
      journal = {\apjl},
         year = 2025,
        month = may,
       volume = {985},
       number = {1},
          eid = {L12},
        pages = {L12},
          doi = {10.3847/2041-8213/add33a},
archivePrefix = {arXiv},
       eprint = {2504.20663},
 primaryClass = {astro-ph.SR},
       adsurl = {https://ui.adsabs.harvard.edu/abs/2025ApJ...985L..12G}
}

@ARTICLE{2023ApJ...943...24P,
       author = {{Panesar}, Navdeep K. and {Hansteen}, Viggo H. and {Tiwari}, Sanjiv K. and {Cheung}, Mark C.~M. and {Berghmans}, David and {M{\"u}ller}, Daniel},
        title = "{Solar Orbiter and SDO Observations, and a Bifrost Magnetohydrodynamic Simulation of Small-scale Coronal Jets}",
      journal = {\apj},
         year = 2023,
        month = jan,
       volume = {943},
       number = {1},
          eid = {24},
        pages = {24},
          doi = {10.3847/1538-4357/aca1c1},
archivePrefix = {arXiv},
       eprint = {2211.06529},
 primaryClass = {astro-ph.SR},
       adsurl = {https://ui.adsabs.harvard.edu/abs/2023ApJ...943...24P}
}

@ARTICLE{2025A&A...694A..71C,
       author = {{Chitta}, L.~P. and {Huang}, Z. and {D'Amicis}, R. and {Calchetti}, D. and {Zhukov}, A.~N. and {Kraaikamp}, E. and {Verbeeck}, C. and {Aznar Cuadrado}, R. and {Hirzberger}, J. and {Berghmans}, D. and {Horbury}, T.~S. and {Solanki}, S.~K. and {Owen}, C.~J. and {Harra}, L. and {Peter}, H. and {Sch{\"u}hle}, U. and {Teriaca}, L. and {Louarn}, P. and {Livi}, S. and {Giunta}, A.~S. and {Hassler}, D.~M. and {Wang}, Y. -M.},
        title = "{Coronal hole picoflare jets are progenitors of both fast and Alfv{\'e}nic slow solar wind}",
      journal = {\aap},
         year = 2025,
        month = feb,
       volume = {694},
          eid = {A71},
        pages = {A71},
          doi = {10.1051/0004-6361/202452737},
archivePrefix = {arXiv},
       eprint = {2411.16513},
 primaryClass = {astro-ph.SR},
       adsurl = {https://ui.adsabs.harvard.edu/abs/2025A&A...694A..71C}
}

@ARTICLE{2024ApJ...963....4S,
       author = {{Sterling}, Alphonse C. and {Panesar}, Navdeep K. and {Moore}, Ronald L.},
        title = "{How Small-scale Jetlike Solar Events from Miniature Flux Rope Eruptions Might Produce the Solar Wind}",
      journal = {\apj},
         year = 2024,
        month = mar,
       volume = {963},
       number = {1},
          eid = {4},
        pages = {4},
          doi = {10.3847/1538-4357/ad1d5f},
archivePrefix = {arXiv},
       eprint = {2401.09560},
 primaryClass = {astro-ph.SR},
       adsurl = {https://ui.adsabs.harvard.edu/abs/2024ApJ...963....4S}
}

@ARTICLE{2014SoPh..289.2733D,
       author = {{De Pontieu}, B. and {Title}, A.~M. and {Lemen}, J.~R. and {Kushner}, G.~D. and {Akin}, D.~J. and {Allard}, B. and {Berger}, T. and {Boerner}, P. and {Cheung}, M. and {Chou}, C. and {Drake}, J.~F. and {Duncan}, D.~W. and {Freeland}, S. and {Heyman}, G.~F. and {Hoffman}, C. and {Hurlburt}, N.~E. and {Lindgren}, R.~W. and {Mathur}, D. and {Rehse}, R. and {Sabolish}, D. and {Seguin}, R. and {Schrijver}, C.~J. and {Tarbell}, T.~D. and {W{\"u}lser}, J. -P. and {Wolfson}, C.~J. and {Yanari}, C. and {Mudge}, J. and {Nguyen-Phuc}, N. and {Timmons}, R. and {van Bezooijen}, R. and {Weingrod}, I. and {Brookner}, R. and {Butcher}, G. and {Dougherty}, B. and {Eder}, J. and {Knagenhjelm}, V. and {Larsen}, S. and {Mansir}, D. and {Phan}, L. and {Boyle}, P. and {Cheimets}, P.~N. and {DeLuca}, E.~E. and {Golub}, L. and {Gates}, R. and {Hertz}, E. and {McKillop}, S. and {Park}, S. and {Perry}, T. and {Podgorski}, W.~A. and {Reeves}, K. and {Saar}, S. and {Testa}, P. and {Tian}, H. and {Weber}, M. and {Dunn}, C. and {Eccles}, S. and {Jaeggli}, S.~A. and {Kankelborg}, C.~C. and {Mashburn}, K. and {Pust}, N. and {Springer}, L. and {Carvalho}, R. and {Kleint}, L. and {Marmie}, J. and {Mazmanian}, E. and {Pereira}, T.~M.~D. and {Sawyer}, S. and {Strong}, J. and {Worden}, S.~P. and {Carlsson}, M. and {Hansteen}, V.~H. and {Leenaarts}, J. and {Wiesmann}, M. and {Aloise}, J. and {Chu}, K. -C. and {Bush}, R.~I. and {Scherrer}, P.~H. and {Brekke}, P. and {Martinez-Sykora}, J. and {Lites}, B.~W. and {McIntosh}, S.~W. and {Uitenbroek}, H. and {Okamoto}, T.~J. and {Gummin}, M.~A. and {Auker}, G. and {Jerram}, P. and {Pool}, P. and {Waltham}, N.},
        title = "{The Interface Region Imaging Spectrograph (IRIS)}",
      journal = {\solphys},
         year = 2014,
        month = jul,
       volume = {289},
       number = {7},
        pages = {2733-2779},
          doi = {10.1007/s11207-014-0485-y},
archivePrefix = {arXiv},
       eprint = {1401.2491},
 primaryClass = {astro-ph.SR},
       adsurl = {https://ui.adsabs.harvard.edu/abs/2014SoPh..289.2733D}
}

@ARTICLE{2011A&A...531A.154G,
       author = {{Gudiksen}, B.~V. and {Carlsson}, M. and {Hansteen}, V.~H. and {Hayek}, W. and {Leenaarts}, J. and {Mart{\'\i}nez-Sykora}, J.},
        title = "{The stellar atmosphere simulation code Bifrost. Code description and validation}",
      journal = {\aap},
         year = 2011,
        month = jul,
       volume = {531},
          eid = {A154},
        pages = {A154},
          doi = {10.1051/0004-6361/201116520},
archivePrefix = {arXiv},
       eprint = {1105.6306},
 primaryClass = {astro-ph.SR},
       adsurl = {https://ui.adsabs.harvard.edu/abs/2011A&A...531A.154G}
}

@article{idl2000,
author = {Stern, Benjamin},
year = {2000},
month = {02},
pages = {1011-},
title = {Interactive Data Language},
doi = {10.1061/40479(204)125}
}

@ARTICLE{2019ApJ...887L...8P,
       author = {{Panesar}, Navdeep K. and {Sterling}, Alphonse C. and {Moore}, Ronald L. and {Winebarger}, Amy R. and {Tiwari}, Sanjiv K. and {Savage}, Sabrina L. and {Golub}, Leon E. and {Rachmeler}, Laurel A. and {Kobayashi}, Ken and {Brooks}, David H. and {Cirtain}, Jonathan W. and {De Pontieu}, Bart and {McKenzie}, David E. and {Morton}, Richard J. and {Peter}, Hardi and {Testa}, Paola and {Walsh}, Robert W. and {Warren}, Harry P.},
        title = "{Hi-C 2.1 Observations of Jetlet-like Events at Edges of Solar Magnetic Network Lanes}",
      journal = {\apjl},
         year = 2019,
        month = dec,
       volume = {887},
       number = {1},
          eid = {L8},
        pages = {L8},
          doi = {10.3847/2041-8213/ab594a},
archivePrefix = {arXiv},
       eprint = {1911.02331},
 primaryClass = {astro-ph.SR},
       adsurl = {https://ui.adsabs.harvard.edu/abs/2019ApJ...887L...8P}
}

@ARTICLE{1978ApJ...226..674F,
       author = {{Feldman}, U. and {Doschek}, G.~A. and {Mariska}, J.~T. and {Bhatia}, A.~K. and {Mason}, H.~E.},
        title = "{Electron densities in the solar corona from density-sensitive line ratios in the N I isoelectronic sequence.}",
      journal = {\apj},
         year = 1978,
        month = dec,
       volume = {226},
        pages = {674-678},
          doi = {10.1086/156649},
       adsurl = {https://ui.adsabs.harvard.edu/abs/1978ApJ...226..674F}
}

@ARTICLE{2007PASJ...59S.655D,
       author = {{de Pontieu}, Bart and {McIntosh}, Scott and {Hansteen}, Viggo H. and {Carlsson}, Mats and {Schrijver}, Carolus J. and {Tarbell}, Theodore D. and {Title}, Alan M. and {Shine}, Richard A. and {Suematsu}, Yoshinori and {Tsuneta}, Saku and {Katsukawa}, Yukio and {Ichimoto}, Kiyoshi and {Shimizu}, Toshifumi and {Nagata}, Shin'ichi},
        title = "{A Tale of Two Spicules: The Impact of Spicules on the Magnetic Chromosphere}",
      journal = {\pasj},
         year = 2007,
        month = nov,
       volume = {59},
        pages = {S655},
          doi = {10.1093/pasj/59.sp3.S655},
archivePrefix = {arXiv},
       eprint = {0710.2934},
 primaryClass = {astro-ph},
       adsurl = {https://ui.adsabs.harvard.edu/abs/2007PASJ...59S.655D}
}

@ARTICLE{2012ApJ...759...18P,
       author = {{Pereira}, Tiago M.~D. and {De Pontieu}, Bart and {Carlsson}, Mats},
        title = "{Quantifying Spicules}",
      journal = {\apj},
         year = 2012,
        month = nov,
       volume = {759},
       number = {1},
          eid = {18},
        pages = {18},
          doi = {10.1088/0004-637X/759/1/18},
archivePrefix = {arXiv},
       eprint = {1208.4404},
 primaryClass = {astro-ph.SR},
       adsurl = {https://ui.adsabs.harvard.edu/abs/2012ApJ...759...18P}
}

@ARTICLE{2018ApJ...854...92T,
       author = {{Tian}, Hui and {Yurchyshyn}, Vasyl and {Peter}, Hardi and {Solanki}, Sami K. and {Young}, Peter R. and {Ni}, Lei and {Cao}, Wenda and {Ji}, Kaifan and {Zhu}, Yingjie and {Zhang}, Jingwen and {Samanta}, Tanmoy and {Song}, Yongliang and {He}, Jiansen and {Wang}, Linghua and {Chen}, Yajie},
        title = "{Frequently Occurring Reconnection Jets from Sunspot Light Bridges}",
      journal = {\apj},
         year = 2018,
        month = feb,
       volume = {854},
       number = {2},
          eid = {92},
        pages = {92},
          doi = {10.3847/1538-4357/aaa89d},
archivePrefix = {arXiv},
       eprint = {1801.06802},
 primaryClass = {astro-ph.SR},
       adsurl = {https://ui.adsabs.harvard.edu/abs/2018ApJ...854...92T}
}

@ARTICLE{2014Sci...346A.315T,
       author = {{Tian}, H. and {DeLuca}, E.~E. and {Cranmer}, S.~R. and {De Pontieu}, B. and {Peter}, H. and {Mart{\'\i}nez-Sykora}, J. and {Golub}, L. and {McKillop}, S. and {Reeves}, K.~K. and {Miralles}, M.~P. and {McCauley}, P. and {Saar}, S. and {Testa}, P. and {Weber}, M. and {Murphy}, N. and {Lemen}, J. and {Title}, A. and {Boerner}, P. and {Hurlburt}, N. and {Tarbell}, T.~D. and {Wuelser}, J.~P. and {Kleint}, L. and {Kankelborg}, C. and {Jaeggli}, S. and {Carlsson}, M. and {Hansteen}, V. and {McIntosh}, S.~W.},
        title = "{Prevalence of small-scale jets from the networks of the solar transition region and chromosphere}",
      journal = {Science},
         year = 2014,
        month = oct,
       volume = {346},
       number = {6207},
          eid = {1255711},
        pages = {1255711},
          doi = {10.1126/science.1255711},
archivePrefix = {arXiv},
       eprint = {1410.6143},
 primaryClass = {astro-ph.SR},
       adsurl = {https://ui.adsabs.harvard.edu/abs/2014Sci...346A.315T}
}

@ARTICLE{2010A&A...519A..49H,
       author = {{He}, J. -S. and {Marsch}, E. and {Curdt}, W. and {Tian}, H. and {Tu}, C. -Y. and {Xia}, L. -D. and {Kamio}, S.},
        title = "{Magnetic and spectroscopic properties of supergranular-scale coronal jets and erupting loops in a polar coronal hole}",
      journal = {\aap},
         year = 2010,
        month = sep,
       volume = {519},
          eid = {A49},
        pages = {A49},
          doi = {10.1051/0004-6361/201014709},
       adsurl = {https://ui.adsabs.harvard.edu/abs/2010A&A...519A..49H}
}

@ARTICLE{2016SSRv..201....1R,
       author = {{Raouafi}, N.~E. and {Patsourakos}, S. and {Pariat}, E. and {Young}, P.~R. and {Sterling}, A.~C. and {Savcheva}, A. and {Shimojo}, M. and {Moreno-Insertis}, F. and {DeVore}, C.~R. and {Archontis}, V. and {T{\"o}r{\"o}k}, T. and {Mason}, H. and {Curdt}, W. and {Meyer}, K. and {Dalmasse}, K. and {Matsui}, Y.},
        title = "{Solar Coronal Jets: Observations, Theory, and Modeling}",
      journal = {\ssr},
         year = 2016,
        month = nov,
       volume = {201},
       number = {1-4},
        pages = {1-53},
          doi = {10.1007/s11214-016-0260-5},
archivePrefix = {arXiv},
       eprint = {1607.02108},
 primaryClass = {astro-ph.SR},
       adsurl = {https://ui.adsabs.harvard.edu/abs/2016SSRv..201....1R}
}

@ARTICLE{2022ApJ...940...85S,
       author = {{Sterling}, Alphonse C. and {Schwanitz}, Conrad and {Harra}, Louise K. and {Raouafi}, Nour E. and {Panesar}, Navdeep K. and {Moore}, Ronald L.},
        title = "{Inconspicuous Solar Polar Coronal X-Ray Jets as the Source of Conspicuous Hinode/EUV Imaging Spectrometer Doppler Outflows}",
      journal = {\apj},
         year = 2022,
        month = nov,
       volume = {940},
       number = {1},
          eid = {85},
        pages = {85},
          doi = {10.3847/1538-4357/ac9960},
archivePrefix = {arXiv},
       eprint = {2210.09233},
 primaryClass = {astro-ph.SR},
       adsurl = {https://ui.adsabs.harvard.edu/abs/2022ApJ...940...85S}
}

@ARTICLE{2015ApJ...815L..16S,
       author = {{Samanta}, Tanmoy and {Pant}, Vaibhav and {Banerjee}, Dipankar},
        title = "{Propagating Disturbances in the Solar Corona and Spicular Connection}",
      journal = {\apjl},
         year = 2015,
        month = dec,
       volume = {815},
       number = {1},
          eid = {L16},
        pages = {L16},
          doi = {10.1088/2041-8205/815/1/L16},
archivePrefix = {arXiv},
       eprint = {1511.07354},
 primaryClass = {astro-ph.SR},
       adsurl = {https://ui.adsabs.harvard.edu/abs/2015ApJ...815L..16S}
}

@ARTICLE{2025ApJ...985L..47B,
       author = {{Bura}, Annu and {Samanta}, Tanmoy and {Prasad}, Avijeet and {Moore}, Ronald L. and {Sterling}, Alphonse C. and {Yurchyshyn}, Vasyl and {Surya}, Arun},
        title = "{Formation of Chromospheric Fan-shaped Jets through Magnetic Reconnection}",
      journal = {\apjl},
         year = 2025,
        month = jun,
       volume = {985},
       number = {2},
          eid = {L47},
        pages = {L47},
          doi = {10.3847/2041-8213/add340},
archivePrefix = {arXiv},
       eprint = {2504.17931},
 primaryClass = {astro-ph.SR},
       adsurl = {https://ui.adsabs.harvard.edu/abs/2025ApJ...985L..47B}
}

@ARTICLE{2016SoPh..291.1129N,
       author = {{Narang}, Nancy and {Arbacher}, Rebecca T. and {Tian}, Hui and {Banerjee}, Dipankar and {Cranmer}, Steven R. and {DeLuca}, Ed E. and {McKillop}, Sean},
        title = "{Statistical Study of Network Jets Observed in the Solar Transition Region: a Comparison Between Coronal Holes and Quiet-Sun Regions}",
      journal = {\solphys},
         year = 2016,
        month = apr,
       volume = {291},
       number = {4},
        pages = {1129-1142},
          doi = {10.1007/s11207-016-0886-1},
archivePrefix = {arXiv},
       eprint = {1604.06295},
 primaryClass = {astro-ph.SR},
       adsurl = {https://ui.adsabs.harvard.edu/abs/2016SoPh..291.1129N}
}
\bibliographystyle{aasjournalv7}
\end{document}